\documentclass[a4paper,preprintnumbers,floatfix,superscriptaddress,prl,twocolumn,showpacs,notitlepage,longbibliography]{revtex4-1}
\usepackage[utf8]{inputenc}
\usepackage[T1]{fontenc}
\usepackage[sc,osf]{mathpazo}
\usepackage{amsmath, amsthm, amssymb,amsfonts,mathbbol,amstext}
\usepackage{graphicx}
\usepackage{dcolumn}
\usepackage{bm}
\usepackage{bbm}
\usepackage{hyperref}
\usepackage{mathtools}
\usepackage{comment}
\usepackage{color}
\usepackage{multirow}

\def\1{\mathbf{1}}
\def\0{\mathbf{0}}

\newcommand{\ket}[1]{| #1 \rangle}
\newcommand{\bra}[1]{\langle #1 |}

\newcommand{\mean}[1]{\left\langle #1 \right\rangle}

\providecommand{\openone}{\mathbbm{1}}
\renewcommand{\rho}{\varrho}

\newcommand{\processnext}[1]{%
  \ifx\listfinish#1\empty\else\listact{#1}\expandafter\processnext\fi}

{\begin{framed}\begin{small}}
{\end{small}\end{framed}}

\DeclareGraphicsExtensions{.pdf,.png,.jpg}

\makeindex

\begin{document}
\title{Quantum Violation of an Instrumental Test}
\date{\today}

\author{Rafael Chaves}
\email{rchaves@iip.ufrn.br}
\affiliation{International Institute of Physics, Federal University of Rio Grande do Norte, 59078-970, P. O. Box 1613, Natal, Brazil}

\author{Gonzalo Carvacho}
\affiliation{Dipartimento di Fisica - Sapienza Universit\`{a} di Roma, P.le Aldo Moro 5, I-00185 Roma, Italy}

\author{Iris Agresti}
\affiliation{Dipartimento di Fisica - Sapienza Universit\`{a} di Roma, P.le Aldo Moro 5, I-00185 Roma, Italy}

\author{Valerio Di Giulio}
\affiliation{Dipartimento di Fisica - Sapienza Universit\`{a} di Roma, P.le Aldo Moro 5, I-00185 Roma, Italy}

\author{Leandro Aolita}
\affiliation{Instituto de F\'{i}sica, Universidade Federal do Rio de Janeiro, Caixa Postal 68528, Rio de Janeiro, RJ 21941-972, Brazil}

\author{Sandro Giacomini}
\affiliation{Dipartimento di Fisica - Sapienza Universit\`{a} di Roma, P.le Aldo Moro 5, I-00185 Roma, Italy}

\author{Fabio Sciarrino}
\email{fabio.sciarrino@uniroma1.it}
\affiliation{Dipartimento di Fisica - Sapienza Universit\`{a} di Roma, P.le Aldo Moro 5, I-00185 Roma, Italy}

\begin{abstract}
\textbf{Inferring causal relations from experimental observations is of primal importance in science. Instrumental tests provide an essential tool for that aim, as they allow to estimate causal dependencies even in the presence of unobserved common causes. In view of Bell's theorem, which implies that quantum mechanics is incompatible with our most basic notions of causality, it is of utmost importance to understand whether and how paradigmatic causal tools obtained in a classical setting can be carried over to the quantum realm. Here we show that quantum effects imply radically different predictions in the instrumental scenario. Among other results, we show that an instrumental test can be violated by entangled quantum states. Furthermore, we demonstrate such violation using a photonic setup with active feed-forward of information, thus providing an experimental proof of this new form of non-classical behavior. Our findings have fundamental implications in causal inference and may also lead to new applications of quantum technologies.}
\end{abstract}
\maketitle

Instrumental variables were originally invented to estimate parameters in econometric models of supply and demand \cite{wright1928tariff} and since then have found a wide range of applications in various other fields \cite{angrist1996identification,Greenland2000}. Remarkably, an instrument allows one to estimate the strength of causal influences between two variables solely from observed data \cite{Balke1997,Pearl2009}, without any assumptions on the functional dependence among them. This is the approach known in quantum information science as "device-independent" \cite{Brunner2014}. For that, an instrumental test is crucial, since it provides empirically testable inequalities
allowing one to check whether one has a valid instrument \cite{pearl1995testability}.

Instrumental inequalities as well as the estimation of causal dependencies are  derived from classical notions of cause and effect that, since Bell's theorem \cite{Bell1964}, we know cannot be taken for granted in quantum phenomena. Given this mismatch between classical and quantum predictions, it is natural to ask how fundamental tools in causal inference behave in a quantum scenario. This has motivated the emerging field of quantum causal modeling \cite{Leifer2013,Fritz2014,Procopio2015,Rubino2017,Henson2014,Chaves2015a,Piennar2015,Costa2016,allen2016quantum}, which has provided sophisticated generalizations of the classical theory of causality \cite{Pearl2009} to the quantum realm, thereby discovering, for example, exciting quantum advantages for causal inference \cite{Fitzsimons2013,Ried15,maclean2016quantum}. Within this new framework, it was shown \cite{Henson2014} that a paradigmatic class of instrumental inequalities \cite{pearl1995testability} are satisfied by quantum mechanics. However, it is not known whether other instrumental inequalities may admit quantum violations. Moreover, even if a given observed statistics is compatible with a classical instrumental causal model, it may well still be the case that quantum effects do offer some sort of enhancement.

In this article, we show that the quantum predictions for the instrumental scenario are radically different from those of classical causality theory. Firstly, we show that a standard measure of causation -- the average causal effect (ACE) \cite{Balke1997,Pearl2009,Schafer2008} -- can be largely over estimated if the latent common cause is a quantum state, a result with both fundamental and applied implications in causal inference. Secondly, we show that, in spite of the results in \cite{Henson2014}, an instrumental inequality can indeed be violated by quantum correlations. Thirdly, we experimentally observe this quantum violation using a photonic set-up with two qubits entangled in polarization equipped with active feed-forward of information \cite{sciarrino_feed_teleportation,minimal_disturbance_Sciarrino}. From a fundamental perspective, our results imply that non-classical behavior emerge even in a very simple causal structure, conceptually different from the paradigmatic Bell scenario \cite{Bell1964} and that in fact leads to a new form of non-classicality stronger than Bell nonlocality.
From an applied viewpoint, our results open an unexplored avenue with potential for new applications of quantum effects, in connection with both causal modeling and information processing.

\begin{figure}[t!]
\center
\includegraphics[width=0.9\columnwidth]{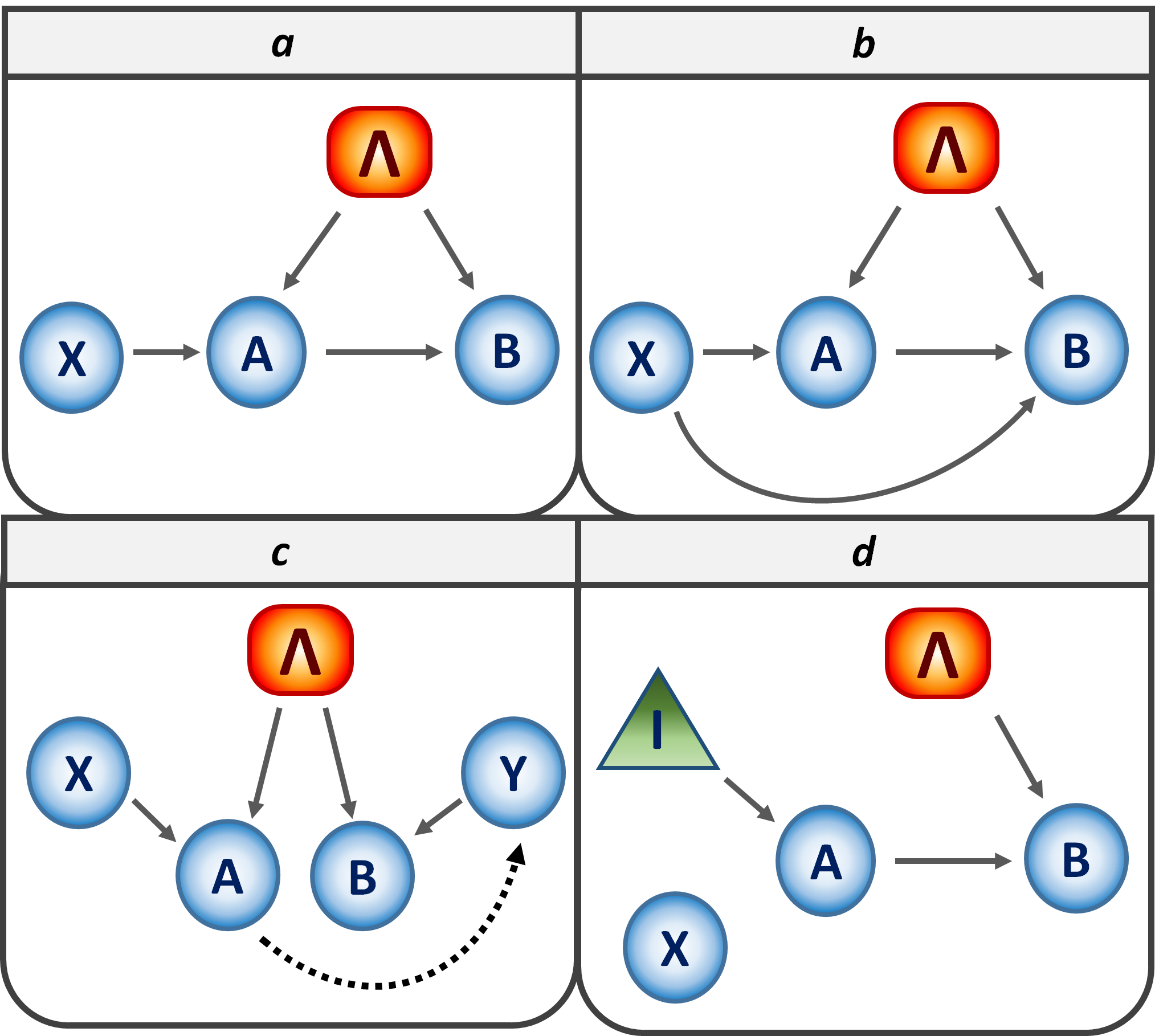}
\caption{ \textbf{DAG representation of causal structures:} a) The instrumental scenario, where $X$ stands for the instrument, $A$ and $B$ are the variables for which we want want to estimate causal influences and $\Lambda$ represent any latent factor correlating them. b) A causal relaxation of the instrumental DAG where direct influences between the instrument $X$ and the ``recovery'' variable $B$ are allowed. c) The bipartite Bell scenario is similar to the instrumental one, however, with no direct causation between $A$ and $B$ and a fourth observable variable $Y$ (acting a ``second instrument'') has to be introduced. The instrumental scenario is equivalent to a non-local causal model allowing for direct causation between $A$ and $Y$ (the dotted arrow). d) Illustration of the effect of an intervention on $A$ making it under influence of a variable $I$ controlled by the experimenter \cite{Costa2016}.}
\label{fig:IDAG}
\end{figure}

An instrument is a random variable $X$, controlled by the experimenter, and satisfying two causal assumptions (see Fig. \ref{fig:IDAG}a). First, it is assumed to be independent of any latent factors (represented by a hidden variable $\Lambda$) that may influence the variables $A$ and $B$ between which we want to infer a causal relation. Second, while $X$ has a direct causal influence over $A$ it does not over $B$; all correlations between $X$ and $B$ are mediated by $A$. As a paradigmatic example, suppose $A$ and $B$ are linearly related by $b=\gamma \cdot a + \lambda$. It then follows that $\gamma= Cov(X,B)/Cov(X,A)$, where $Cov(X,A)=\mean{X,A}-\mean{X}\mean{A}$ is the covariance between $A$ and $X$, and similarly for $Cov(X,B)$. Hence, from the observed correlations between the instrument and the variables of interest we can estimate the strength ($\gamma$) of the causal influence connecting them, even without any information about $\Lambda$. Moreover, as already mentioned, an instrument allows for causal estimation even without any such functional-dependence knowledge \cite{Pearl2009,Balke1997}; that is, in the device-independent scenario. To this end, however, we first need to introduce a general framework for causal inference that relies solely on observed probabilities.

The causal relationships between $n$ random variables $(X_1,\dots, X_n)$ can be graphically described by directed acyclic graphs (DAGs), examples of which are shown in Fig. \ref{fig:IDAG}. Each node in the graph represents a variable, and each directed edge a causal relation between two variables \cite{Pearl2009}. Every variable can be expressed as a deterministic function $x_i = f_i(pa_i, u_i)$ of its graph-theoretic parents $PA_i$ and a local noise term $U_i$, implying that the probability $p(\mathbf{x}) = p(x_1, . . . , x_n)$ has a Markov decomposition as $p(\mathbf{x})=\prod_{i=1}^{n}p(x_i \vert pa_{i})$. A central concept in the theory of causality is that of an intervention \cite{Pearl2009,Costa2016,allen2016quantum} defined as the act of forcing a given variable, say $X_i$, to assume a value $x_i$. This operation is denoted as $\mathrm{do}(X_i=x^{\prime}_i)=\mathrm{do}(x^{\prime}_i)$. The effect of an intervention is to replace the original mechanism $x_i = f_i(pa_i, u_i)$ by $x_i=x^{\prime}_i$, while keeping all other functionals $f_{j \neq i}$ unchanged. That is, an intervention erases all incoming arrows to given variable (see Fig. \ref{fig:IDAG}d).

The importance of interventions stems from the fact that they allow one to distinguish correlations due to common causes from those due to direct causation (even in the absence of an instrumental variable). In fact, interventions are at the core of the definition of causal influence. For instance, if $A$ has a direct causal influence over $B$, then typically $p(b \vert \mathrm{do}(a)) \neq p(b \vert \mathrm{do}(a^{\prime}))$ (for some $a\neq a^{\prime}$). In contrast, if the correlations are solely due to the common ancestor $\Lambda$, then $p(b \vert \mathrm{do}(a)) = p(b \vert \mathrm{do}(a^{\prime}))$ (for all $a$ and $a^{\prime}$). These considerations lead to a widely used measure of causation called average causal effect (ACE) \cite{Pearl2009,Schafer2008}, defined by
\begin{equation}
\label{ACE}
\mathrm{ACE}_{A \rightarrow B} = \sup_{a,a^{\prime},b} \vert p(b \vert \mathrm{do}(a))-p(b \vert \mathrm{do}(a^{\prime})) \vert.
\end{equation}
The ACE can be understood as the maximum observable change in the distribution of $B$ caused by changes on $A$, on average (over all  hidden common factors grouped into $\Lambda$). However, for practical, fundamental or even ethical (as in randomized clinical trials) reasons, one frequently does not have access to interventions. In that case, one must rely only on observational data -- that is, on Bayesian conditionals $p(b \vert a)$ rather than do-conditionals $p(b \vert \mathrm{do}(a))$--  to infer causations. It is there where instruments come into play.

%%%%%%%%%%%%%%%%%%%%%%%%%%%%%

To illustrate, suppose $A$ represents the compliance of patients to a treatment assigned (that is, whether they actually follow the treatment) and $B$ stands for their recovery, while $X$ stands for a randomized (thus independent of $\Lambda$) treatment assignment (for example, a random choice between true drug versus a placebo without letting the patience know about the choice). There might be several factors -- for example, social, economical or personal-- affecting jointly both the patient's compliance and the recovery that are not directly observable and act as a hidden source of correlations. Thus, in practice, all the empirical data available from the instrumental test is contained in the probability distribution $p(a,b,x)$ of the observable variables. Moreover, since the instrumental variable $X$ is under the experimenter's control, all the relevant empirical information is encoded in the conditional distribution $p(a,b \vert x)$. The ACE is lower bounded by a linear function of this distribution. For instance, if all variables are binary ($x=1,2$ and $a,b=0,1$), it follows that \cite{Pearl2009,Balke1997}
\begin{eqnarray}
\label{ACEbound}
\mathrm{ACE}_{A \rightarrow B} \geq & & \,2\,p(a=0,b=0\vert x=1)-2\\ \nonumber
& &+p(a=1,b=1\vert x=1)+p(b=1\vert x=2).
\end{eqnarray}
That is, remarkably, the instrument $X$ allows one to estimate the effect of hypothetical interventions on $A$ even if they are not actually performed. 

This reasoning, however, relies on a classical description, implying that the distribution decomposes as
\begin{equation}
\label{instrumental_deco}
p(a,b \vert x )=\sum_{\lambda} p(a \vert x, \lambda)p(b \vert a,\lambda)p(\lambda),
\end{equation}
where we have simply used the causal assumptions underlying the classical instrumental causal model, namely that $p(\lambda \vert x)= p(\lambda)$ and $p(b \vert a,x, \lambda)=p(b \vert a, \lambda)$. However, this is not the most general form of correlations when the nodes are granted quantum capabilities. The full quantization of the instrumental causal structure, where all nodes represent quantum systems, can be studied with the frameworks of \cite{Costa2016,allen2016quantum}. Here, for concreteness, we consider that all three observable variables, $X$, $A$ and $B$, are still classical but that the latent variable $\Lambda$ is now a quantum system. This quantum model produces observable correlations of the form
\begin{equation}
\label{eq:prob}
p_{\mathrm{Q}}(a,b \vert x )=\mathrm{Tr} \left[ (M^{x}_{a} \otimes M^{a}_{b}) \rho \right],
\end{equation}
where $M^{x}_{a}$ describes a measurement operator depending on the measurement choice $x$ with outcome $a$ (similarly to $ M^{a}_{b}$) and $\rho$ is the density operator that plays the quantum analogue of the classical hidden variable $\Lambda$.

Our first result is to show that the quantum prediction for $\mathrm{ACE}_{A \rightarrow B}$ is in stark contrast with the classical one, such that, in particular, Eq. \eqref{ACEbound} no longer holds for quantum correlations of the form of equation \eqref{eq:prob}. To see this, let $\Lambda$ be a two-qubit system in the state 
$
\rho=v\,\ket{\phi^{+}}\bra{\phi^{+}}+(1-v)\,\openone/4,
$
where $\openone/4$ the maximally mixed state and $\ket{\phi^{+}}=(\ket{\uparrow\uparrow}+\ket{\downarrow \downarrow})/\sqrt{2}$ is the maximally entangled one, and $0\leq v \leq 1$. Let $M^{x}_{a}$ and $M^{a}_{b}$ correspond to the von Neumann projectors corresponding to the measurements outcomes $a$ and $b$ of the observables $O_{A}^{x}$ and $O_{B}^{a}$, respectively, where $O_{A}^{x=0}=\sigma_Z$, $O_{A}^{x=1}=\sigma_X$, $O_{B}^{a=0}=-\sin{(\pi/8)}\,\sigma_X+\cos{(\pi/8)}\,\sigma_Z$, and $O_{B}^{a=1}=(\sigma_X+\sigma_Z)/\sqrt{2}$, where $\sigma_{X,Z}$ are Pauli matrices and $\ket{\uparrow}, \ket{\downarrow}$ are the eigenstates of $\sigma_{z}$. The resulting quantum correlations are compatible with classical instrumental models -- that is, admit a decomposition of the form \eqref{instrumental_deco}. Therefore, Eq. \eqref{ACEbound} can be applied leading to $\mathrm{ACE}_{A \rightarrow B}\gtrsim 0.91\,v-0.75$, which is strictly greater than zero for all $v \gtrsim 0.82$. However, surprisingly, $\mathrm{ACE}_{A \rightarrow B}$ becomes identically zero in the quantum setting.

Analogously to the classical case (see Methods), we have that 
$
p_{\mathrm{Q}}(b \vert \mathrm{do}(a))=\mathrm{Tr}\left[M^{\mathrm{do}(a)}_{b}\, \rho_B \right],
$
where $\rho_B$ is the reduced state of $\rho$ over the qubit measured at node $B$. 
Hence, the generalization of ACE to the quantum instrumental model, $\mathrm{QACE}_{A \rightarrow B}$, is given by
\begin{equation}
\label{QACE}
\mathrm{QACE}_{A \rightarrow B}=\sup_{a,a^{\prime},b} \left\vert \mathrm{Tr}\left[\left(M^{\mathrm{do}(a)}_{b}-M^{\mathrm{do}(a^{\prime})}_{b}\right)\, \rho_B \right]\right\vert
\end{equation}
For the particular $\rho$ under consideration, it follows that $\rho_B=\mathrm{Tr}_A\left[ \rho \right]=\frac{\openone}{2}$, implying that $p_{\mathrm{Q}}(b \vert \mathrm{do}(a))=1/2$ for all $b$ and $\mathrm{do}(a)$, so that one automatically gets $\mathrm{QACE}_{A \rightarrow B}=0$.
Clearly, any bipartite state with a maximally mixed reduced state over $B$ subject to arbitrary von Neumann measurements will, by the same arguments, produce quantum instrumental correlations with null QACE. The fact that, the quantum ACE can be lower than the classical one can be seen as a quantum enhancement in terms of the causal strength required to reproduce some data, in the sense that the average statistics observable at $B$ must be more sensitive to interventions in $A$ if the causal model is classical. Interventions are device-dependent operations, since they rely on precise control of the physical mechanisms underlying the observable statistics \cite{Ried15,Ringbauer2016,Costa2016}. A natural question is then whether one can observe discrepancies between the classical and quantum predictions in a device-independent manner and, in particular, without the need of interventions. In what follows, we answer this question affirmatively by finding a new instrumental inequality with quantum violation.
 
In the scenario with dichotomic variables, the only non-trivial class of inequalities has been characterized in ref. \cite{pearl1995testability} and cannot be violated by quantum instrumental correlations \cite{Henson2014}.
However, allowing the instrumental variable to take $3$ possible values $x=1,2,3$ while keeping the measurement outcome variables binary ($a,b=0,1$) is already enough to find quantum violations. As described in the Supplementary Information, the classical correlations compatible with the instrumental DAG define a convex set that can be characterized by standard tools in convex optimization \cite{boyd2004convex}. For the scenario considered, apart from the known class without quantum violations, a new class of non-trivial instrumental inequalities appear \cite{bonet2001instrumentality,Popescu1994}, represented by
\begin{equation}
\mathcal{I}=
-\mean{B}_1+2\mean{B}_2+\mean{A}_1-\mean{AB}_1+2\mean{AB}_3 \leq 3,
\label{new_instrumental}
\end{equation}
where $\mean{AB}_x= \sum_{a,b=0,1} (-1)^{a+b} p(a,b \vert x)$, and similarly for $\mean{A}_x$ and $\mean{B}_x$.

This inequality can be violated by projective measurements on any pure entangled state of two qubits (see Supplementary Information), obtaining the maximal violation with a maximally entangled state. Specifically, for the state $\ket{\phi^{+}}$, maximal violation can be achieved measuring observables $O_{A}^{x=1}=-(\sigma_X+\sigma_Z)/\sqrt{2}$, $O_{A}^{x=2}=\sigma_X$ and $O_{A}^{x=3}=\sigma_Z$ and $O_{B}^{a=0}=(\sigma_X+\sigma_Z)/\sqrt{2}$ and $O_{B}^{a=1}=(\sigma_Z-\sigma_X)/\sqrt{2}$. This leads, through Eq. \eqref{eq:prob}, to quantum correlations that give the value
\begin{equation}
\mathcal{I}_{\mathrm{Q}}=1+2\sqrt{2}
\end{equation}
for the left-hand side of Eq. \eqref{new_instrumental}, thus violating the inequality. Further, we can also ask what is the maximal violation of Eq. \eqref{new_instrumental}, if, instead of quantum correlations, we now allow for post-quantum (non-signalling) correlations \cite{Popescu1994}. As shown in the Supplementary Information, the maximal value of the left-hand side in this case is $\mathcal{I}_{\mathrm{NS}}=5$, achieved if the underlying correlation is given by a generalization of the paradigmatic Popescu-Rohrlich-box \cite{Popescu1994}. 

From the classical perspective, the violation of an instrumental inequality can only be due to the fact that the underlying dynamics is not described by an instrumental causal model \cite{pearl1995testability}. For instance, referring again to the example of randomized clinical trials, due to the well known placebo effect, the patient's expectation can cause the placebo to have a similar effect (mediated by the patient's compliance) to the true treatment. However, if the patient realizes that the assigned treatment is a placebo pill instead of the true drug, the placebo effect might be gone: independently of whether the patient complies to follow the assigned treatment, this will be useless for the recovery. That is, the treatment X generates a direct (not mediated by the compliance $A$) causal influences on the recovery $B$, as in Fig. 1b. As a result, $X$ is clearly no longer an instrumental variable. Using a measure $\mathcal{C}_{X \rightarrow B}$ to quantify the causal influence between $X$ and $B$ (see Methods) we can show that
\begin{equation}
\label{eq:Relaxation}
\min \mathcal{C}_{X \rightarrow B} = \max{ \left[ \frac{\mathcal{I}-3}{4},0 \right]}.
\end{equation}
This gives a quantitative interpretation to any violation of Eq.\eqref{new_instrumental} in terms of the minimum causal influence $X \rightarrow B$ required to explain the observed data with a classical causal model. In contrast, the novel instrumental inequality of Eq.\eqref{new_instrumental} can be violated by quantum instrumental causal models of the type presented above, which clearly satisfy the two causal assumptions defining the instrumental scenario.

\begin{figure*}[!t]
\includegraphics[width=1.8\columnwidth]{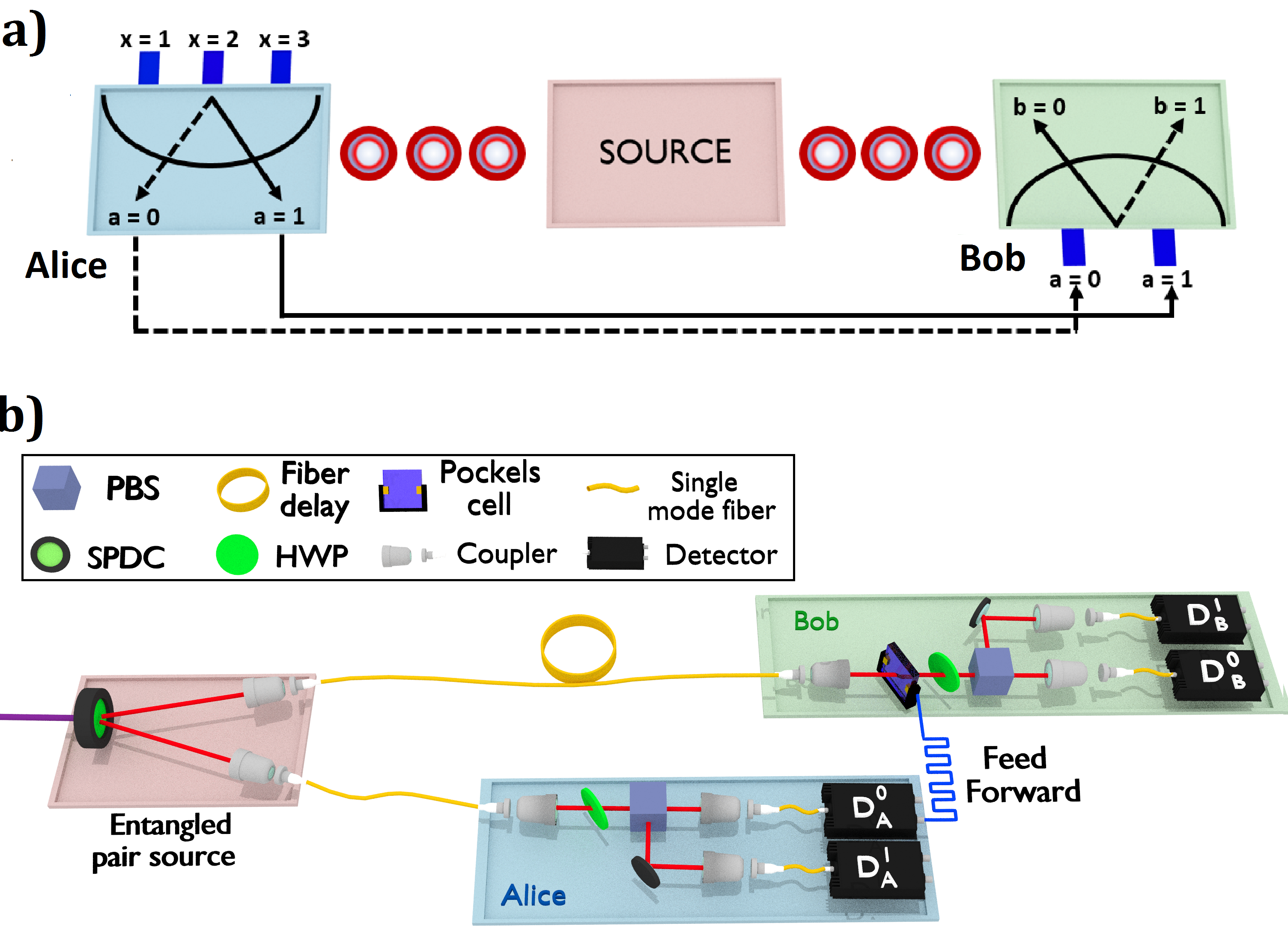}
\caption{ \textbf{Experimental apparatus for the violation of the instrumental inequality}: a) A pictorial and device-independent representation of the experiment using ``black-boxes'' (blue buttons standing for measurement choices and the meters indicating the measurement outcome). The violation of inequality equation \eqref{new_instrumental} does not depend on any assumptions about the internal working of the measurement apparatus and solely relies on the observed probability $p(a,b \vert x)$. b)  A polarization-entangled photon pair is generated via Spontaneous Parametric Down-Conversion (SPDC) in a nonlinear crystal. One photon (qubit $A$) is sent to the Alice's station, where one of the three observables ($O^{x=1}$, $O^{x=2}$ and $O^{x=3}$) is measured via the rotation of motorized half-wave plate ($HWP_A$) followed by a polarizing beam splitter (PBS). Detector $D_{A}^{0}$ acts as trigger for the application of a 1500 V voltage on the Pockels cell, whenever the measurement output $0$ is registered. The second photon (qubit B) is delayed 720 ns before arriving to Bob's station by employing a single-mode fiber 158.6 m long. After leaving the fiber the photon passes through the Pockels cell, followed by a $HWP_B$ fixed at $-11.25^{\circ}$ and a PBS. If the Pockels cell has been triggered (in case of A measurement outcome is $0$), its action combined to the fixed $HWP_B$ allows us to project onto $\frac{\sigma_{z}+\sigma_{x}}{\sqrt{2}}$. Otherwise (if $A$ measurement outcome is $1$), the Pockels cell acts as the identity and we project onto $\frac{\sigma_{z}-\sigma_{x}}{\sqrt{2}}$.}
\label{fig:ExpSetupV1}
\end{figure*}

To experimentally implement the instrumental causal structure shown in Fig. \ref{fig:IDAG}a, we exploited the photonic platform depicted in Fig.\ref{fig:ExpSetupV1}. Through a process of spontaneous parametric down-conversion, we generated the maximally entangled photonic state $\ket{\phi^{+}}$.
The entangled photons are sent on two different paths and the measurement outcomes of the polarization of each photon constitute the variables $A$ and $B$. We therefore will refer to the two polarizations as the $A$ (Alice's) and $B$ (Bob's) qubits. The instrumental causal structure requires direct causation between Alice and Bob, since the choice of the measurement to be performed on Bob's qubit must depend on the outcome registered by Alice.
Therefore, we need an experimental apparatus whose time for a single measurement run allows the measurement and registration of Alice's outcome, the communication between the two parties and then the application of the selected measurement operator on qubit B.  
To achieve that with an active exchange of information from Alice to Bob, we implemented an active feed-forward scenario by employing a Pockels cell. In this way, our experimental apparatus relies only on the application of a voltage to switch between the two measurement settings for Bob.

The first photon (qubit $A$) is sent to Alice's station in order to project onto the eigenspaces of the three different observables $O_{A}^{x=1,2,3}$. This is done through a rotated half wave plate (HWP) followed by a polarizing beamsplitter (PBS), which enables us to perform all of the necessary measurements, of the form $\cos{\left(4\theta\right)}\sigma_{z}+\sin{\left(4\theta\right)}\sigma_{x}$, where $\theta$ is the orientation angle of the optical axis of the waveplate with respect to the PBS optical axis. After the PBS, a photon detector is placed at each of its two output modes ($D_A^{1}$, $D_A^{0}$). Each time a photon is detected by $D_A^{0}$ the electrical signal is split in two, sending one to the coincidence counter, while the other is used as a trigger for the action of the Pockels cell used in the measurement of the second photon.

For qubit $B$ the measurement apparatus is made of the Pockels cell, followed by a HWP, characterized by a rotation angle of $-11.25^{\circ}$, and a PBS. When no voltage is applied to the cell, it acts as the identity; otherwise it works as a HWP in its optical axis performing a unitary evolution given by the $\sigma_{z}$ operator. This enables us to perform projective measurements on the eigenvectors of the observable $\frac{\sigma_{z}-\sigma_{x}}{\sqrt{2}}$ when no voltage is applied to the cell and of $\frac{\sigma_{z}+\sigma_{x}}{\sqrt{2}}$ otherwise.
After the PBS, two photon detectors ($D_B^{0}$, $D_B^{1}$) are placed, one for each output mode. To achieve the active feed-forward of information, the second photon is delayed by sending it into a long fiber, thus allowing the measurement on the first qubit to be performed.

\begin{figure}[t!]
\center
\includegraphics[width=1.0\columnwidth]{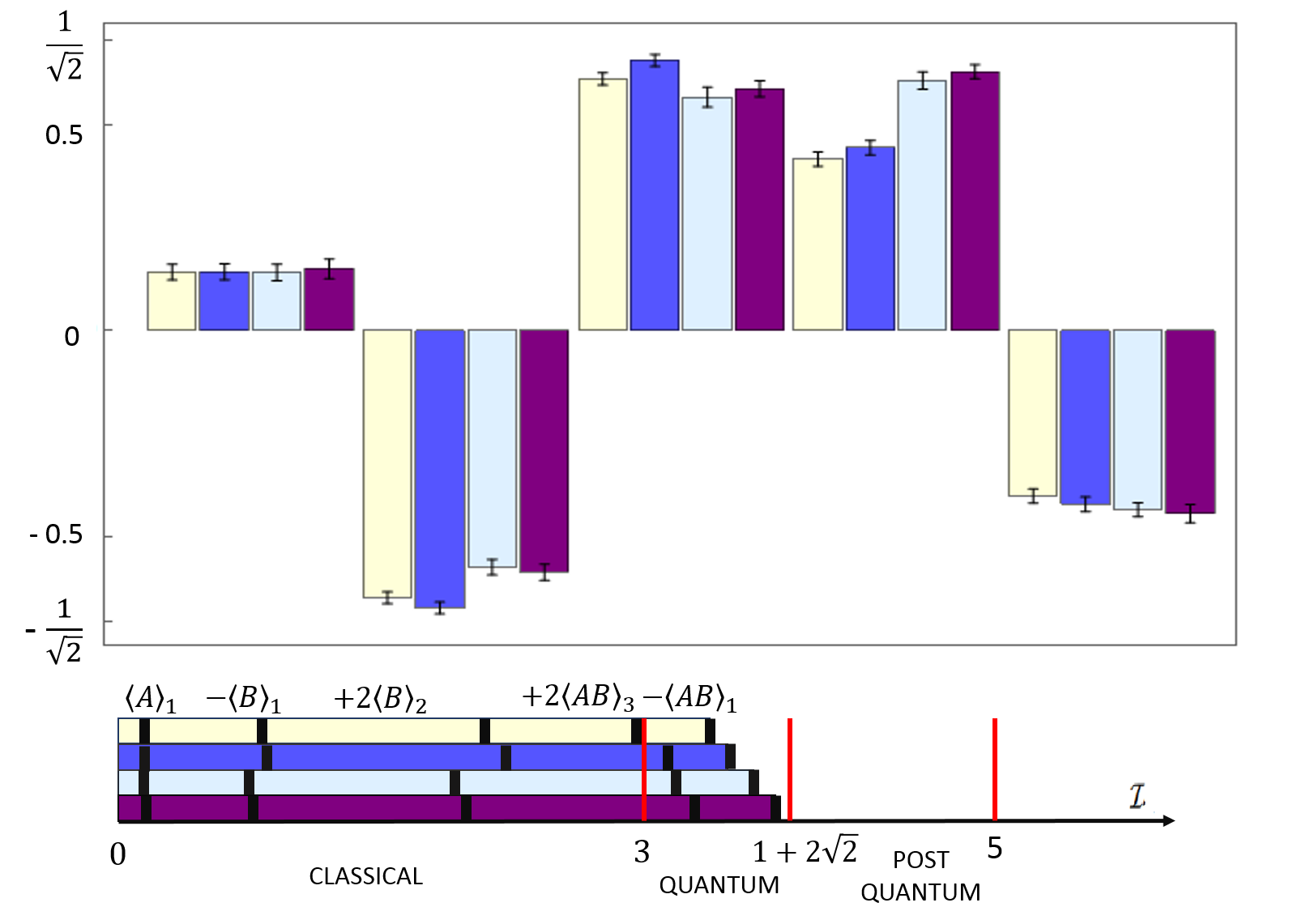}
\caption{\textbf{Experimental results for the violation of the instrumental inequality:} in the top part, the experimental results for the five expectation values in inequality \eqref{new_instrumental}. The four different colours represent the experimental values obtained in the following configurations: with active feed-forward without accidental counts correction (yellow), with active feed-forward and accidental counts correction (blue); without active feed-forward with raw data (light blue) and without active feed-forward with accidental counts correction (purple) (using post-selection of the events, see Methods). The five sets of columns represent respectively the obtained values for $\mean{A}_1$, $\mean{B}_1$, $\mean{B}_2$, $\mean{AB}_3$ and $\mean{AB}_1$. In the bottom part, the length of the three coloured horizontal bars show the final results for the inequality values: they all lie in the range of a quantum violation. The error bars indicate the intervals $\pm$3 $\sigma$ of the 5 terms involved in the inequality and obtained as weighted averages over various sets of experimental data. For each set of collected data, the uncertainty of the number of photon coincidences was estimated by Monte Carlo simulations.}
\label{fig:example}
\end{figure}

A coincidence counter, synchronizing the four detectors, distinguishes the signals generated by entangled photons from accidental counts. Our corresponding experimental violation of equation \eqref{new_instrumental} with active feed-forward yields $\mathcal{I}=3.258 \pm 0.020$, surpassing the classical limit by $12.9$ standard deviations with raw data and $\mathcal{I}=3.358 \pm 0.020$ with accidental counts correction (see Fig.\ref{fig:example}). This value is fully compatible with our theoretical model which considers the noise in the adopted state and the efficiency of the Pockels cell, as detailed in the Supplementary Information. The causal relaxations defined in eq. (\ref{eq:Relaxation}) needed to classically explain the experimental violations are: $\min \mathcal{C}_{X \rightarrow B} = \left(6.45\pm0.50\right)\times10^{-2}$ and $\min\mathcal{C}_{X \rightarrow B} =\left (8.95\pm0.50\right)\times10^{-2}$ without accidental counts. We also implemented the DAG structure simulating the communication channel between A and B, resorting to the post-selection of the experimental data, obtaining a higher violation. As it turns out, the instrumental scenario can be understood as a usual Bell scenario where one performs a post-selection of data where $y=a$. Alternatively, the instrumental inequality can be seen as a constraint bounding a non-local hidden variable model with measurement dependence (see Fig. \ref{fig:IDAG}c and Methods). Thus, its violation also proves a stronger form of non-classicality as compared with Bell's theorem, since the latter refer to local hidden variable models. Note that, in the post-selection experiment, we measure the observable $O^{a=1}$ even in the case where the measurement outcome is $a=0$ (and vice-versa) and such information is irrelevant for the evaluation of instrumental inequality. This way, the implementation of the active feed-forward leads to an overall experimental speed-up.

In conclusion, we have shown that classical and quantum predictions for the instrumental scenario differ radically. As a consequence, basic results from causal inference need to be reevaluated and reinterpreted. Firstly, when compared with quantum theory, classical causality concepts can lead to a significant overestimation of the average causal effect, a key notion for causal inference and, in fact, a primal application of instrumental variables \cite{Pearl2009,Schafer2008}. This can also be seen as a quantum advantage in terms of the strength of causal influences required to reproduce  observations, and leads us, in addition, to a natural definition of a quantum average causal effect. Secondly, we have shown that an instrumental inequality can be violated by quantum instrumental causal models physically implementable by local measurements on a quantum state augmented with classical communication of one of the measurement outcomes. Since these models satisfy the two causal assumptions defining the instrumental scenario, the usual interpretation of violations of instrumental inequalities as revealing incompatibility with the instrumental causal structure no longer holds.

Thirdly, using a photonic setup with active feed-forward, we have experimentally implemented the instrumental causal structure. We observed a quantum violation of an instrumental inequality using maximally entangled states encoded in the polarization of two photons. Recent experiments \cite{Ringbauer2016,Carvacho2017,Saunders2017,Ringbauer2017} have explored causal structures beyond those of Bell's theorem; however, to our knowledge, no other experiment in this context has been performed using active transmission of information, which adds whole new challenges in itself. We highlight that, just as in Bell setups, our experiment is also subjected to loopholes. Namely, first, we are not subjected the locality loophole of Bell experiments, but one has to rely on the assumption that all the information from the setting $X$ to the outcome $B$ flows through the outcome $A$ (see Supplementary Information for a detailed discussion). Second, as discussed also in the Supplementary Information, to avoid the detection efficiency loophole we would need total efficiencies above $90\%$, meaning that, in practice, we still have to rely on the fair sampling assumption. 

From a fundamental perspective, our results show that non-classical behaviors can emerge in quantum experiments with very simple causal structures other than those of Bell scenarios. 
The instrumental scenario is indeed very much related to a Bell scenario, quoting Judea Pearl, one of the pioneers in the causality theory \cite{Pearl2009}: \emph{``The similarity of the instrumental inequality to Bell's inequality in quantum physics is not accidental... The instrumental inequality can, in a sense, be viewed as a generalization of Bell's inequality for cases where direct causal connection is permitted to operate between the correlated observables, A and B.''} 
In view of our results, instrumental inequalities and Bell inequalities are much more connected than that: instrumental inequalities can be understood as Bell inequalities characterizing non-local hidden variable models with measurement dependence. Thus, the quantum violation of an instrumental inequality brings to light a new form of non-classicality stronger than Bell nonlocality in its usual form. Furthermore, this quantum effect can be given a quantitative interpretation in terms of the amount of causal relaxations in the instrumental causal structure for a classical model to reproduce the quantum predictions.

To end up with, we believe that this work opens several avenues for future research. On an applied side, it is natural to ask whether results usually associated with Bell scenarios, such as device-independent cryptography \cite{Vazirani2014}, randomness generation \cite{Pironio2010,colbeck2011,Colbeck2012,GMTDAA13,BRGHHSW16} and amplification \cite{Colbeck2012,GMTDAA13,BRGHHSW16}, or self-testing \cite{Mayers2003}, can also be extended to the instrumental case. Also, in view of the results  of refs. \cite{Fitzsimons2013,Ried15}, which show that quantum correlations can allow for causal inference with  observational data in situations where classical correlations require interventions, it is interesting to explore similar possibilities in the instrumental scenario. 
Besides, as for the newly introduced causality measure, the quantum average causal effect, it would be desirable to find non-trivial lower bounds in terms of observable data alone (without quantum interventions), analogous to equation \eqref{ACEbound} for the classical average causal effect.
Finally, at a more basic level, it would be highly desirable to understand the different kinds of non-classical behaviors emerging from imposing causal structures of growing complexity to quantum systems in general.
The answer to these questions could bring new insights on the nature of quantum theory itself and, as a consequence, on its advantages and limitations for information processing.

\section{acknowledgments}
RC and LA acknowledge financial support from the Brazilian ministries MEC and MCTIC. In addition, LA is also grateful to the Brazilian agencies CAPES, CNPq, FAPERJ, and INCT-IQ for financial support. This work was supported by the ERC-Starting Grant 3D-QUEST (3D-Quantum Integrated Optical Simulation; grant agreement number 307783): http://www.3dquest.eu. QUCHIP-Quantum Simulation on a Photonic Chip grant agreement number 641039. G. C. thanks Becas Chile and Conicyt for a doctoral fellowship

\section{Additional Information}
Supplementary information is available in the online version of the paper.

\subsection{Author contributions}
G.C., I.A, V.D.G, S.G and F.S. devised and performed the experiment; R.C. and L.A. developed the theoretical tools; all the authors discussed the results and contributed to the writing of the manuscript.
\subsection{Competing financial interests}
The authors declare no competing financial interest.

\bigskip
\section{Methods}
\subsection{Experimental Details}

Photon pairs were generated in a parametric down conversion source, composed by a nonlinear crystal beta barium borate (BBO) of 2 mm-thick injected by a pulsed pump field with $ \lambda = 392.5$ nm. After spectral filtering and walk-off compensation, photons of $ \lambda = 785$ nm are sent to the two measurement stations A and B.
The crystal used to implement active feed-forward is a $\mathrm{LiNbO}_{3}$ high-voltage micro Pockels Cell made by Shangai Institute of Ceramics with <1 ns risetime and a fast electronic circuit transforming each Si-avalanche photodetection signal into a calibrated fast pulse in the kV range needed to activate the Pockels Cell is fully described in \cite{sciarrino_feed_teleportation}. To achieve the active feed-forward of information, the photon sent to Bob's station needs to be delayed, thus allowing the measurement on the first qubit to be performed. The amount of delay was evaluated considering the velocity of the signal transmission through a single mode fiber and the activation time of the Pockels cell. We have used a fiber 158.6 m long, coupled at the end into a single mode fiber that allows a delay of 720 ns of the second photon with respect to the first.

\subsection{Violation of the instrumental inequality by post-selection of the data}

It is also possible to obtain a violation of the equation \eqref{new_instrumental}  without active feed-forward but via post-selection. That is, one performs all the measurements of the observables in A and B, without the voltage application to the Pockels cell on the B path, and then post-selects the relevant combinations of outcomes by Alice and observables by Bob at the end. The observables in station B can be therefore measured with the Pockels cell constantly working as the identity operator and varying the angle of the HWP after the Pockels cell. After performing the six possible combinations of observables, we can simulate a classical channel of communication between the parties by post-selecting only the events appearing in the inequality equation \eqref{new_instrumental}. The maximum value reached in our experimental set-up under these conditions, with the accidental counts correction, was $3.621\pm 0.023$, while in the case of raw data was $3.500 \pm 0.023$ (see Fig.\ref{fig:example}). The former value corresponds to a violation of the inequality equation \eqref{new_instrumental} of almost $27$ sigmas. As expected, this violation increases the instrumental relaxation needed to fit the observed correlations, which according to equation \eqref{eq:Relaxation} yields $\min \mathcal{C}_{X \rightarrow B} = \left(1.552\pm0.058\right)\times10^{-1}$.

\subsection{Instrumental scenario as a non-local hidden variable model with measurement dependence}

The post-selection experiment described above can be understood as a regular Bell test where instead of testing quantum mechanics against local hidden variable model we actually test a stronger non-local hidden variable model also allowing for measurement dependence.

Performing a regular Bell experiment (described by $p(a,b \vert x,y)$) and post-selecting $y=a$ is equivalent in the DAG description to a causal structure with an arrow between the $A$ and $Y$ variables (see Fig. \ref{fig:IDAG}c). In this case we have the Markov decomposition given by
\begin{equation}
p(a,b,x,y,\lambda)=p(a \vert x,\lambda)p(b \vert y,\lambda) p(x)p(y\vert a)p(\lambda)
\end{equation}
implying that
\begin{eqnarray}
p(a,b \vert x) =& &\frac{\sum_{\lambda,y}p(a,b,x,y,\lambda)}{p(x)} \\ \nonumber
& & =\sum_{\lambda,y} p(a \vert x,\lambda)p(b \vert y,\lambda) p(y\vert a)p(\lambda).
\end{eqnarray}
Choosing $y=a$ then we arrive at the decomposition given by
\begin{equation}
p(a,b \vert x)=\sum_{\lambda}p(a\vert x,\lambda)p(b\vert a,\lambda)p(\lambda),
\end{equation}
that is exactly the same as the instrumental decomposition in equation \eqref{instrumental_deco}.
Furthermore, since the variable $A$ has as a parent the hidden variable $\Lambda$, this model allows for correlations between the variables $\Lambda$ with $Y$ ($p(y,\lambda) \neq p(y)p(\lambda)$); that is, it is also a measurement-dependent hidden variable model. An example of how this classical model can be used to simulate maximally nonlocal correlations is given in the Supplementary Information.

\subsection{Quantifying causal influences between $X$ and $B$}

Along similar lines to the causal interpretation of the violation of Bell inequalities \cite{Chaves2015b}, we can quantify the degree of ``non-instrumentality'' of the classical instrumental causal model by how much we have to relax the instrumental assumptions. In the instrumental scenario we have two options: allow for direct correlations between variables $X$ and $B$; allow for correlations between $X$ and $\Lambda$. Both relaxations are reminiscent, respectively, of the locality and measurement independence relaxations considered in Bell's theorem \cite{Chaves2015b,Toner03,Hall2010,Barrett2010}. In the following, we will consider in detail the causal relaxation $X \rightarrow B$ (see Fig. \ref{fig:IDAG}b) discussed  in the clinical-trial analogy above. This can be quantified \cite{Chaves2015b} by a measure similar to the ACE \eqref{ACE}, known as the direct causal influence from $X$ to $B$:
\begin{equation}
\label{CAB}
\mathcal{C}_{X \rightarrow B} = \sup_{x,x^{\prime},a,b} \sum_{\lambda}p(\lambda)\vert p(b\vert \mathrm{do}(x),a,\lambda)-p(b\vert \mathrm{do}(x^{\prime}),a,\lambda) \vert,
\end{equation}
This gives the maximum shift in the probability of $B$ caused by interventions on $X$ (averaged over the latent factors represented by the hidden variable $\Lambda$ and discounted the causal influence mediated by $A$). In particular, note that here, differently from equation \eqref{ACE}, one takes the average over $\lambda$ outside of the modulo. Thus, this measure is zero if and only if $X$ has no direct causal influence over $B$.

The connection between the violation of the instrumentality inequality equation \eqref{new_instrumental} and $\mathcal{C}_{X \rightarrow B}$ as given by equation \eqref{eq:Relaxation} can be proved using a standard approach in convex optimization and is detailed in the Supplementary Information.

\subsection{Interventions in quantum causal models}

To define the quantum generalization of the average causal effect (ACE) we have to first introduce the notion of interventions for quantum causal models \cite{Costa2016}. As in the classical case, the quantum do-conditional probability $p_{\mathrm{Q}}(b \vert \mathrm{do}(a))$ is defined by a modified system dynamics whereby the causal influences of $X$ and $\Lambda$ on $A$ are erased and $A$ is forced to take a value $do(a)$ of our choice, while the causal influences of $A$ and $\Lambda$ on $B$ are kept intact. Applying this prescription to the quantum correlations of equation \eqref{eq:prob} gives 
\begin{align}
\nonumber
p_{\mathrm{Q}}(b \vert \mathrm{do}(a))&=\mathrm{Tr}\left[\openone\otimes M^{\mathrm{do}(a)}_{b}\, \rho \right]\\
\label{eq:q_do_cond}
&=\mathrm{Tr}\left[M^{\mathrm{do}(a)}_{b}\, \rho_B \right],
\end{align}
where $\rho_B$ is the reduced state of $\rho$ over the qubit measured at node $B$. 
Hence, we arrive at \eqref{QACE}. In particular, notice that the fact that $\mathrm{QACE}_{A \rightarrow B}$ depends on  $\rho_B$ (instead  of, e.g., the conditional state for $B$ given the measurement input $x$  and output $a$ at $A$) reflects the fact that only the average statistics observable at $B$ are relevant for the QACE.

\subsection{Data availability}

The data that support the plots within this paper and other findings of this study are available from the corresponding author upon request.

\end{document}

\title{Supplementary Information: Quantum Violation of an Instrumental Test}
\date{\today}

\author{Rafael Chaves}
\email{rchaves@iip.ufrn.br}
\affiliation{International Institute of Physics, Federal University of Rio Grande do Norte, 59078-970, P. O. Box 1613, Natal, Brazil}

\author{Gonzalo Carvacho}
\affiliation{Dipartimento di Fisica - Sapienza Universit\`{a} di Roma, P.le Aldo Moro 5, I-00185 Roma, Italy}

\author{Iris Agresti}
\affiliation{Dipartimento di Fisica - Sapienza Universit\`{a} di Roma, P.le Aldo Moro 5, I-00185 Roma, Italy}

\author{Valerio Di Giulio}
\affiliation{Dipartimento di Fisica - Sapienza Universit\`{a} di Roma, P.le Aldo Moro 5, I-00185 Roma, Italy}

\author{Leandro Aolita}
\affiliation{Instituto de F\'{i}sica, Universidade Federal do Rio de Janeiro, Caixa Postal 68528, Rio de Janeiro, RJ 21941-972, Brazil}

\author{Sandro Giacomini}
\affiliation{Dipartimento di Fisica - Sapienza Universit\`{a} di Roma, P.le Aldo Moro 5, I-00185 Roma, Italy}

\author{Fabio Sciarrino}
\email{fabio.sciarrino@uniroma1.it}
\affiliation{Dipartimento di Fisica - Sapienza Universit\`{a} di Roma, P.le Aldo Moro 5, I-00185 Roma, Italy}

\maketitle

\subsection{Geometric description of the set of classical correlations in the instrumental DAG}
As discussed in the main text, the Markov condition for the instrumental DAG implies that any observable probability distribution compatible with it can be decomposed as 
\begin{equation}
\label{deco_app}
p(a,b \vert x )=\sum_{\lambda} p(a \vert x, \lambda)p(b \vert a,\lambda)p(\lambda).
\end{equation}
As we can see, this decomposition defines a convex set (the convex sum over the probabilities for the variable $\Lambda$). For discrete variables, this set is in fact a polytope, that is, a convex set with finitely many extremal points and that can be equivalently described by finitely many linear inequalities that represent the boundaries of the polytope (from which the non-trivial represent instrumental inequalities). 

As mentioned in the main text, the probabilities appearing in the decomposition above can without loss of generality be expressed as deterministic (response) functions. More precisely, we have that
\begin{eqnarray}
&&a=D_{a}(x,\lambda), \\
&&b=D_{b}(a,\lambda).
\end{eqnarray}
That is, for a given value of $\lambda$ the variable $a$ is a deterministic function of $x$; similarly $b$ is a deterministic function of $a$. We can think of the variable $\lambda$ as selecting which function to use in a given run of the physical process (with probability $p(\lambda)$). The number of possible deterministic functions to choose from will intrinsically depend on the domain of the variables. Namely, the number of deterministic functions for variable $A$ is given by $\vert a \vert ^{\vert x \vert} $ and for variable $B$ is given by $\vert b \vert ^{\vert a \vert}$, where $\vert a \vert$ is the number of values taken by the random variable $A$ (similarly for $B$ and $X$). Thus the total number of deterministic functions is given by $\vert a \vert ^{\vert x \vert} \vert b \vert ^{\vert a \vert}$; each of these corresponding to one of the extremal points defining the convex set defined by \eqref{deco_app}.

Given the list of extremal points of a polytope, we can find its dual description in terms of facets, that is, linear inequalities of the probabilities $p(a,b \vert x)$ that generically can be written as
\begin{equation}
\sum_{a,b,x} c_{a,b,x} p(a,b \vert x) \leq B_{\mathrm{Classical}},
\end{equation}
where $c_{a,b,x}$ are real coefficients and $B_{\mathrm{Classical}}$ corresponds to the maximum values achievable by the extremal points.
Further, this dualization procedure can be performed using standard convex optimization software \cite{PORTA}. A given correlation is compatible with the instrumental DAG if and only if it respects all of these inequalities. If any of them is violated we thus have an unambiguous proof that some of the causal assumptions is incompatible with the correlation under test.

\subsection{Quantum and post-quantum violations of the instrumental inequality}

Our interest is to find the optimal quantum violations of inequality (eq. (6) in the main text)
\begin{equation}
\mathcal{I}=
-\mean{B}_1+2\mean{B}_2+\mean{A}_1-\mean{AB}_1+2\mean{AB}_3 \leq 3,
\label{new_instrumental_SI}.
\end{equation}
That is, to find the optimal quantum state and measurements producing correlations according to
\begin{equation}
\label{PQ}
p_{\mathrm{Q}}(a,b \vert x)=\mathrm{Tr} \left[ (M^{x}_{a} \otimes M^{a}_{b}) \rho \right].
\end{equation}

\begin{figure}[t!]
\center
\includegraphics[width=1\columnwidth]{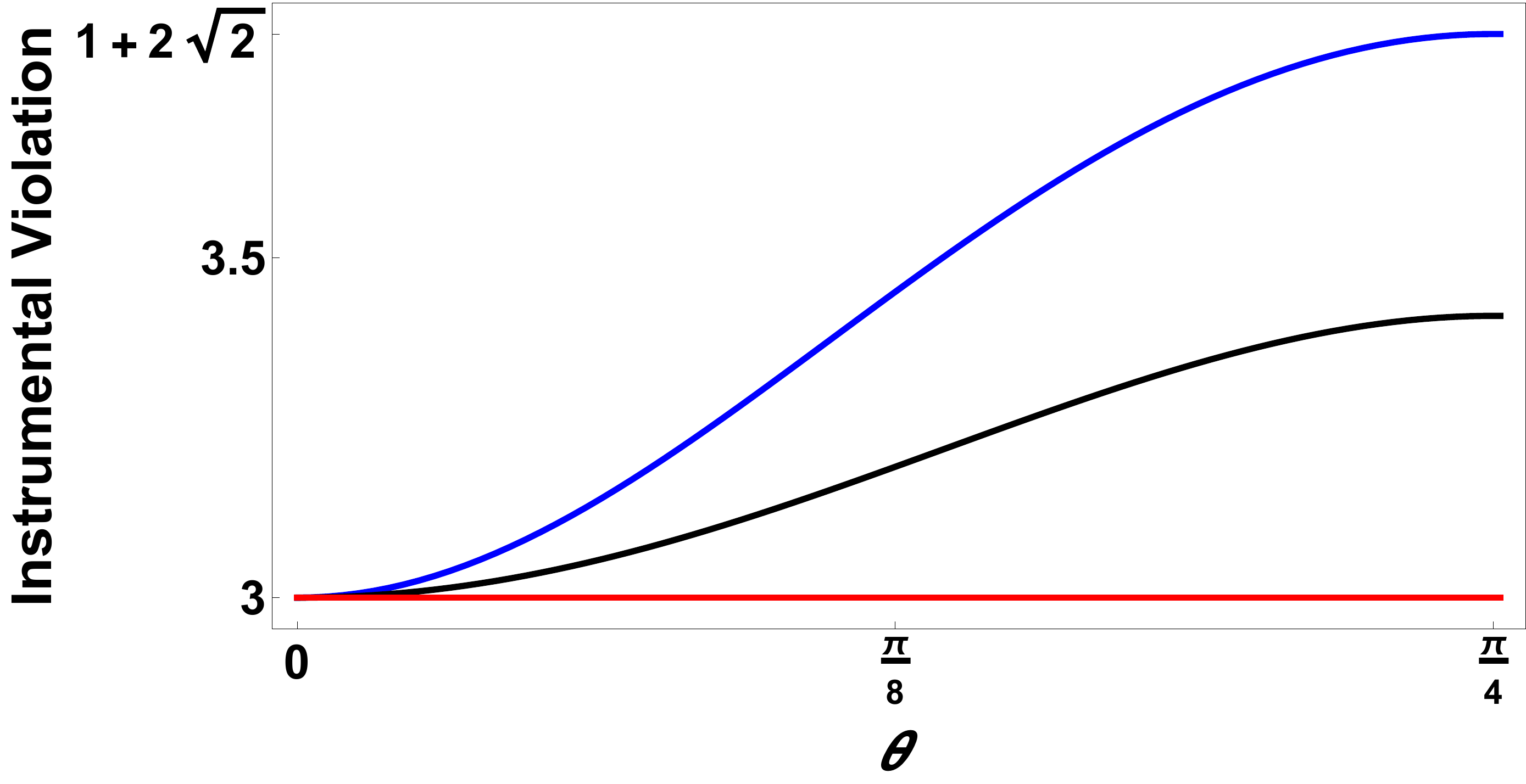}
\caption{\textbf{Violation of the instrumental inequality \eqref{new_instrumental_SI} as function of $\theta$:} every pure entangled state violates the inequality and the maximum is achieved by the maximally entangled state. The blue curve shows the optimal violation obtained numerically by considering projective measurements on both qubits. The black curve shows the (non-optimal) violation following \eqref{non_optimal}. The red line shows the classical limit.}
\label{fig:Vplot}
\end{figure}

To that aim we have restricted our attention to pure two-qubit states of the form $\ket{\Psi}=\cos{(\theta)}\ket{\uparrow \uparrow}+\sin{(\theta)}\ket{\downarrow \downarrow}$ (because of the convexity of the set of quantum correlations we do not need to consider mixed states in the optimization). As for the measurement observables we have considered qubit projective measurements. Thus, the variable $A$ is associated to the measurement outcome of the measurement performed on the first qubit of the pair and, following the prescription of the instrumental DAG, the measurement settings for the second qubit (the measurements outcomes which correspond to the variable $B$) can depend on the measurement outcome of the first measurement. By numerically optimizing (in Mathematica) the measurement settings for each value of $\theta$ we observe that every entangled state ($\theta \neq n \pi/2$) violates the inequality and that the maximal violation is obtained for maximally entangled states ($\theta =\pi/4$) (see Fig. \ref{fig:Vplot}). 

To prove analytically that every entangled pure state of two-qubits (considering without loss of generality $0 < \theta \leq \pi/4$) violates the inequality it is enough to set $O^{x=1}=-(\sigma_X+\sigma_Z)/\sqrt{2}$, $O^{x=2}=\sigma_X$ and $O^{x=3}=\sigma_Z$ for the first qubit (producing the outcome $a$) and $O^{a=0}=(\cos{(\theta)}\sigma_Z+\sin{(\theta)}\sigma_X)$ and $O^{a=1}=\sigma_Z$ for the second qubit (producing the outcome $b$).
The corresponding value for inequality \eqref{new_instrumental_SI} is given by
\begin{equation}
\label{non_optimal}
\frac{1}{4}(4+(8+3\sqrt{2})\cos(\theta)-2\sqrt{2}\cos(2\theta)-\sqrt{2}\cos(3\theta)),
\end{equation}
that can be easily seen to be larger than $3$ for any $0 < \theta \leq \pi/4$ (see Fig. \ref{fig:Vplot}). We notice that this choice of parameters does not lead to the optimal violation for a given $\theta$ but it is enough for the purpose of showing violations.

%%%%%%%%%%%%%%%%%%%%%%%%%%%%%

Instead of considering quantum resources, we can consider what is the maximum violation of the instrumental inequality provided by non-signalling (post-quantum) resources. To that aim, consider a given non-signalling (NS) distribution $p_{\mathrm{NS}}(a,b \vert x,y)$ respecting the NS constraints
\begin{eqnarray}
\nonumber
& & p_{\mathrm{NS}}(a \vert x)= \sum_{b}p_{\mathrm{NS}}(a,b \vert x,y)=\sum_{b}p_{\mathrm{NS}}(a,b \vert x,y^{\prime}), \\  \nonumber
& & p_{\mathrm{NS}}(b \vert y)= \sum_{a}p_{\mathrm{NS}}(a,b \vert x,y)=\sum_{a}p_{\mathrm{NS}}(a,b \vert x^{\prime},y),
\end{eqnarray}
for all $x,x^{\prime},y,y^{\prime}$.

\begin{figure}[t!]
\center
\includegraphics[width=1\columnwidth]{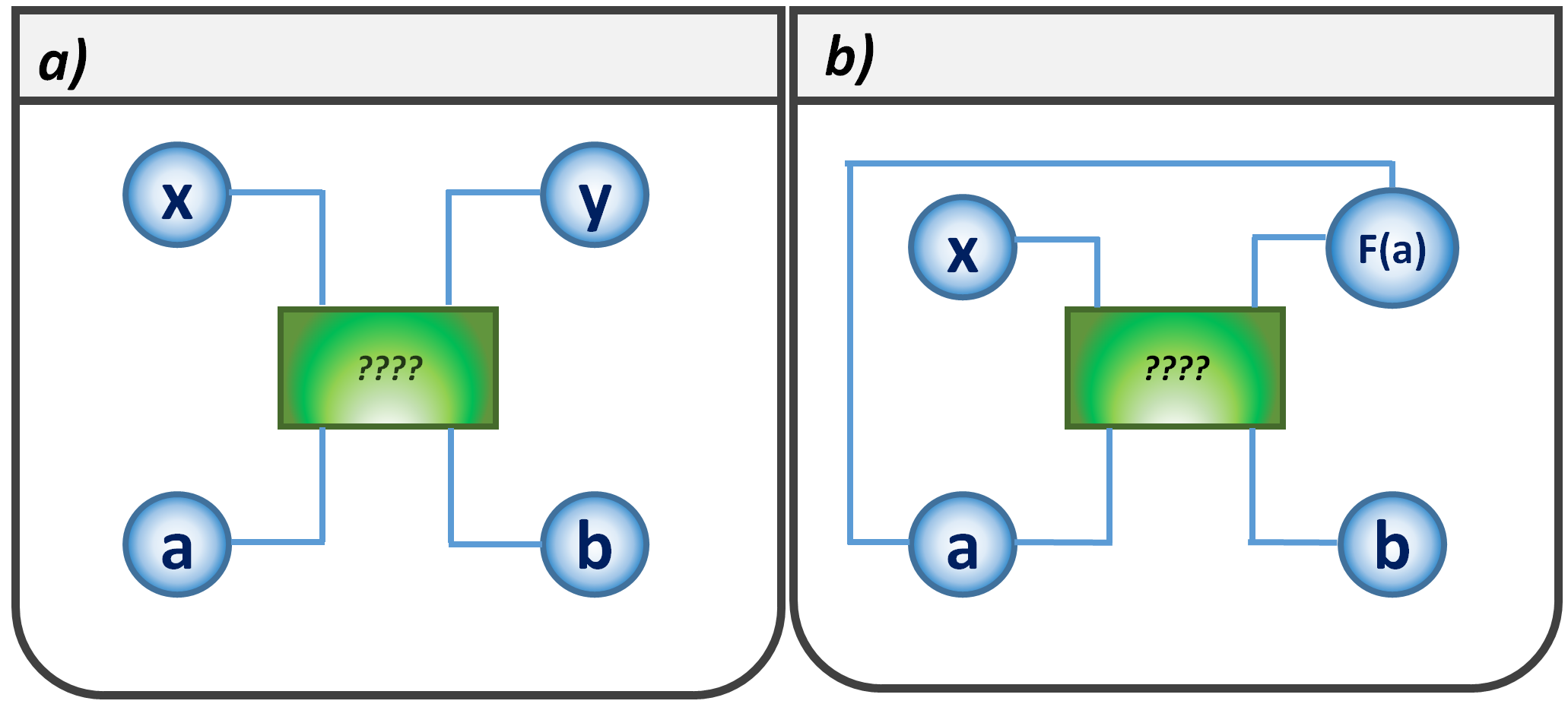}
\caption{\textbf{Black-box representation of a physical process:} a) The black box associated to a probability distribution $p(a,b \vert x,y)$. b) The wired box used to produce the probability $p(a,b \vert x)$.}
\label{fig:Bbox}
\end{figure}

NS correlations can represented by the black-box process illustrated in Fig. \ref{fig:Bbox}. Just as in the classical and quantum cases, one of the inputs of this black-box can depend on the output obtained earlier. That is, in this case we are considering ``wirings'' of probability distributions \cite{Barrett2005}, as illustrated in Fig. \ref{fig:Bbox}. The probability distribution for the instrumental scenario obtainable from NS resources via these wirings is thus given by
\begin{equation}
p(a,b \vert x)= p_{\mathrm{NS}}(a,b \vert x,f(a)),
\end{equation} 
where $f(a)$ is an arbitrary deterministic function with $\vert a \vert$ possible inputs and $\vert a \vert$ possible outputs that describes a wiring from Alice's output to Bob's input. In turn, for a fixed $f(a)$ it is a simple linear program (LP) to find what is the maximum violation of the instrumental inequality \eqref{new_instrumental_SI}.
Since there are finitely many functions $f(a)$ there is also a finite number of LPs one has to solve. Doing that, we obtain that the maximum violation is $\mathcal{I}_{\mathrm{NS}}=5$. The latter is obtained for $f(a)=a$ and with $p_{\mathrm{NS}}(a,b \vert x,y)$ the simple variation of the Popescu-Rohrlich (PR) box \cite{Popescu1994} shown in Table  \ref{tabPR}.

\begin{table}[hp]
$p_{\mathrm{NS}}(a,b \vert x,y)=$
\begin{tabular}{c | c || c c | c c |}
$x\backslash y$ & & \multicolumn{2}{c|}{0} & \multicolumn{2}{c|}{1} \\
\hline
 & $a\backslash b$      &   0            &   1               &   0   &   1   \\
\hline
\hline
\multirow{3}{*}{0} & 0  &   $\frac{1}{2}$                      &   $\frac{1}{2}$            &       $\frac{1}{2}$        &      $\frac{1}{2}$        \\
                   & 1  &   0                      &   0            &       0        &      0        \\
\hline 
\multirow{2}{*}{1} & 0  &   $\frac{1}{2}$               &   0            &       0       &     $\frac{1}{2}$         \\
                   & 1  &   0                      & $\frac{1}{2}$  &      $\frac{1}{2}$         & 0 \\
\hline
\multirow{2}{*}{2} & 0  &   $\frac{1}{2}$               &   0         &       $\frac{1}{2}$        &      0        \\
                   & 1  &   0                     & $\frac{1}{2}$  &      0      & $\frac{1}{2}$ \\
\hline
\end{tabular}
\caption{A matrix representation of the non-signalling distribution maximally violating the instrumental inequality. The rows stand for the $x$ and $a$ values while the columns stand for $y$ and $b$. For $x=0$ we see that $A$ always outputs $a=0$ while B produces a random bit $b$. Restricting to the inputs $x=1,2$ and transforming them as $x=1 \rightarrow x=0$ and $x=2 \rightarrow x=1$ we have a symmetry of the PR-box given by $p(a,b \vert x,y)=\frac{1}{2}\delta_{a \oplus b, (x+1)y}$.}\label{tabPR}
\end{table}

\subsection{Quantifying causal relaxations in the instrumental DAG}
Observed a violation of an instrumental inequality it is natural to consider by how much we have to relax the underlying causal assumptions of the instrumental model in order to explain the observed correlations. As discussed in the main text, the instrumental DAG subsumes two causal assumptions: i) the independence of the instrumental variable $X$ from the hidden variable $\Lambda$ ($p(x,\lambda)=p(x)p(\lambda)$) and ii) the fact that all the correlations between $X$ and $B$ are mediated via $A$ and $\Lambda$ ($p(b\vert a,x,\lambda)=p(b\vert a,\lambda)$).

\begin{figure}[t!]
\center
\includegraphics[width=1\columnwidth]{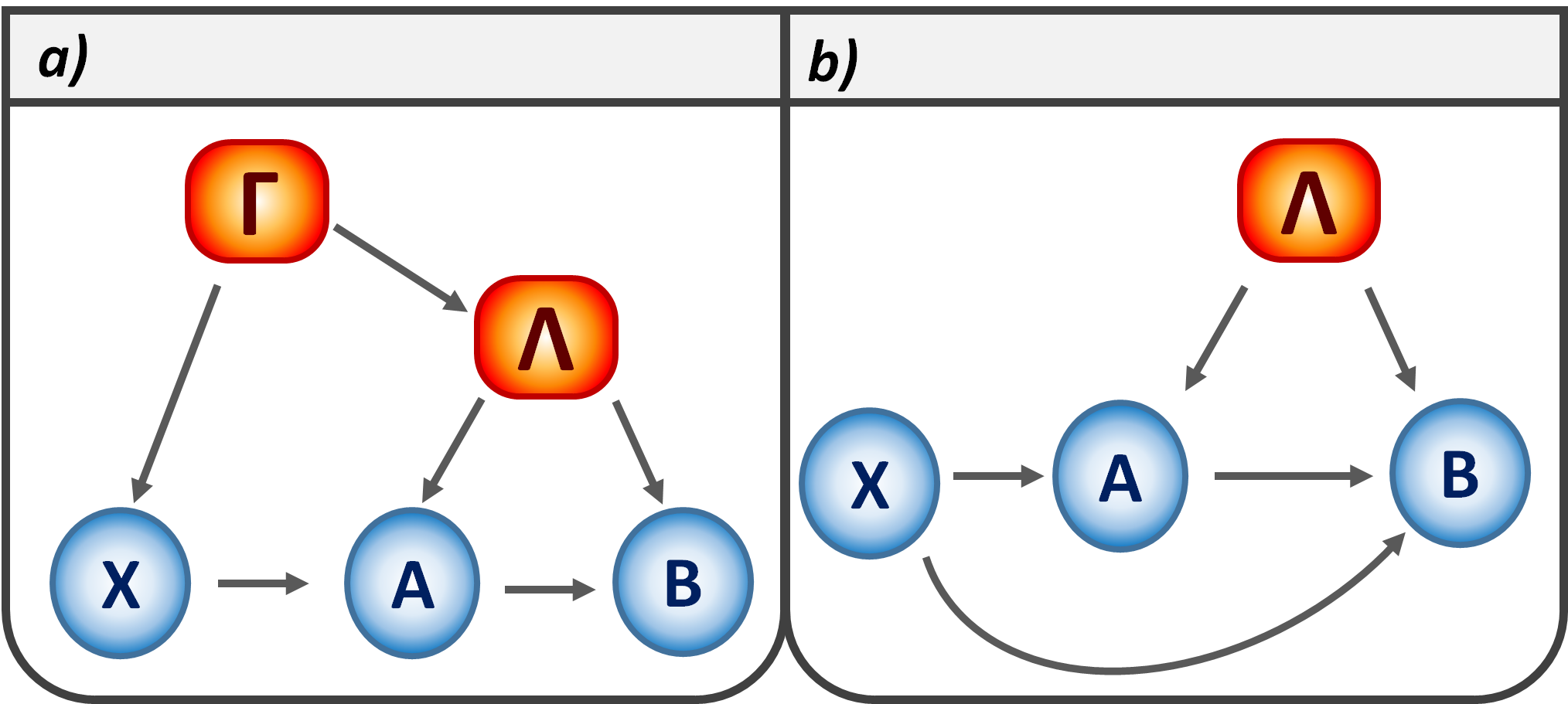}
\caption{\textbf{Causal relaxations of the instrumental DAG:} a) Measurement independence relaxation where the hidden variable $\Lambda$ is allowed to be correlated with the instrumental variables $X$. b) A relaxation where a direct causal influence from $X$ into $B$ is allowed.}
\label{fig:relax}
\end{figure}

The graphical language of DAGs immediately allows us to represent the causal relaxations in both cases (see Fig. \ref{fig:relax}). In the first case, we allow for correlations between $X$ and $\Lambda$ to be mediated by a common ancestor $\Gamma$. In turn, in the second DAG we allow for a direct causal influence (a directed edge) between variables $X$ and $B$. The DAG representation, however, does not specify the strength of the arrows, that is, by how much we have the relax the corresponding causal assumptions. To that aim, we first have to introduce a measure of correlations and/or direct causal influence. Below we introduce the measures we consider and using the techniques developed in \cite{Chaves2015b} show how they can be estimated from given observed correlations via a linear program.

For the first case, the observed probability distribution has a decomposition given by
\begin{eqnarray}
\label{eq:Mdep}
p(a,b \vert x )= & & \sum_{\lambda,\gamma} p(a \vert x, \lambda)p(b \vert a,\lambda)p(\lambda,\gamma \vert x) \\ \nonumber
= & &\frac{1}{p(x)} \sum_{\lambda,\gamma} p(a \vert x, \lambda)p(b \vert a,\lambda)p(\lambda\vert \gamma) p(x \vert \gamma)p(\gamma) .
\end{eqnarray}
Without loss of generality, we model such correlations by introducing an additional hidden variable $\Gamma$ which serves as a common ancestor for $X$, and  $\Lambda$.
This suggests to decompose this common ancestor into $\gamma=(\gamma_x,\gamma_{\lambda})$. We can assume $x=\gamma_x$ and $\lambda=\gamma_{\lambda}$ (that is, they are deterministic functions) without loss of generality. If the observable variables $a$, $b$ and $x$ are discrete variables then finitely many different instances of $\gamma$ suffice to fully characterize the common ancestor's influence.

Alternatively, we can represent \eqref{eq:Mdep} as
\begin{equation}
\p = T \q,
\end{equation}
where  $p(a,b|x)$ is represented by a vector $\p$ with components $\p_j$ labeled by the indexes $j=(a,b,x)$ and the distribution of $\gamma$ is associated with a vector with components $\q_\gamma=p(\Gamma = \gamma)$. In turn, 
$T$ is a matrix with elements $T_{j,\gamma} = \frac{1}{p(x)}\delta_{x,\gamma_x}\delta_{a,D_a(x,\gamma_{\lambda})}
\delta_{b,D_b(a,\gamma_{\lambda})}= \frac{1}{p(x)}\delta_{a,D_a(\gamma_x,\gamma_{\lambda})}
\delta_{b,D_b(a,\gamma_{\lambda})}$.

Since the correlations between $X$ and $\Lambda$ are mediated by a third variable $\Gamma$, it is reasonable to consider as a measure of correlations the total variation distance given by
\begin{equation}
\mathcal{M}_{X:\Lambda}=\sum_{x,\lambda} \vert p(x,\lambda)-p(x)p(\lambda) \vert.
\end{equation}
There are certainly other good measures one can employ, however, the total variation distance has the good advantage of providing a non-trivial lower bound, via the Pinsker inequality \cite{Fedotov2003}, to the mutual information between $X$ and $\Lambda$:
\begin{equation}
I(X:\Lambda) \geq \frac{1}{2}\mathcal{M}_{X:\Lambda}^2\log{\e},
\end{equation}
a quantity that is highly non-linear as a function of the probabilities and thus can in practice only be estimated numerically.

In the second case (the one discussed in the main text), we want to quantify the causal influence of the arrow $X \rightarrow B$, that is, the causal model allows for distributions $p(a,b \vert x)$ of the form
\begin{equation}
p(a,b \vert x )=\sum_{\lambda} p(a \vert x, \lambda)p(b \vert a,x,\lambda)p(\lambda).
\end{equation}
Similarly to what we have done in the case of measurement dependence, we can represent this model as
\begin{equation}
\p = W \q,
\end{equation}
where again  $p(a,b|x)$ is represented by a vector $\p$ with components $\p_j$ labeled by the indexes $j=(a,b,x)$ and the distribution of $\lambda$ is associated with a vector with components $\q_\lambda=p(\Lambda = \lambda)$. In turn, 
$W$ is a matrix with elements $W_{j,\lambda} = \delta_{a,D_a(x,\lambda)}
\delta_{b,D_b(a,x,\lambda)}$.

Since in this case we do have a direct arrow between the variables, it is more appropriate to consider a measure of causal influence rather than a simple measure of correlations. As discussed in the main text, in this case we consider the measure of causal influence defined as
\begin{equation}
\mathcal{C}_{X \rightarrow B} = \sup_{x,x^{\prime},a,b} \sum_{\lambda}p(\lambda)\vert p(b\vert a,do(x),\lambda)-p(b\vert a,{do(x^\prime}),\lambda) \vert.
\end{equation}
Further, we notice that in this case, since the variables $X$ and $B$ have no common ancestors, the intervention (the do operation) is totally equivalent to a simple conditional operation, that is, $p(b \vert do(x))=p(b \vert x)$ and thus 
\begin{equation}
\mathcal{C}_{X \rightarrow B} = \sup_{x,x^{\prime},a,b} \sum_{\lambda}p(\lambda)\vert p(b\vert a,x,\lambda)-p(b\vert a,{x^\prime},\lambda) \vert.
\end{equation}

Given some observed probability distribution or given the violation of some instrumental inequality (represented by a given matrix $V$ such that $V \p$ gives the corresponding inequality), our aim is to estimate the minimum relaxation necessary to explain it. In other terms what is the minimum value of $\mathcal{M}_{X:\Lambda}$ or $\mathcal{C}_{X \rightarrow B}$. That is, we are interested in the following minimization problems:
\begin{eqnarray}
\underset{ {\bf q} \in \RR^{n}}{\minimize} & & \quad  \mathcal{M}_{X:\Lambda} \label{min_1}  \\ \nonumber
\st & & \quad V T{\bf q} = V \p  \\
& & \quad \langle \1_n, {\bf q} \rangle = 1 \\ \nonumber
& & \quad {\bf q}  \geq \0_n.
\end{eqnarray}
or
\begin{eqnarray}
\underset{ {\bf q} \in \RR^{n}}{\minimize} & & \quad  \mathcal{C}_{X \rightarrow B} \label{min_2}  \\ \nonumber
\st & & \quad V W{\bf q} = V \p  \\
& & \quad \langle \1_n, {\bf q} \rangle = 1 \\ \nonumber
& & \quad {\bf q}  \geq \0_n.
\end{eqnarray}
Following the results in \cite{Chaves2015b}, these optimization problems can be casted a linear programs and thus analytical and computationally efficient solutions can be found. Notice that the same approach can be applied to compute the average causal effect $\mathrm{ACE}_{A \rightarrow B}$ and the direct causal influence $\mathcal{C}_{X \rightarrow B}$ (eqs. (1) and (12) of the main text).

\subsection{A simple example showing how post-selection can simulate non-local correlations}

Here we illustrate how in the usual Bell scenario the post-selection of data given by $y=a$ can be used to simulate non-local and signalling correlations through local hidden variable (LHV) models. However, as discussed in the main text such post-selection is allowed if instead of testing quantum mechanics againts LHV models we test it against non-local hidden variable models, more specifically a model where the outcome $A$ has a direct causal influence over $Y$ (thus also allowing for measurement dependence).

Consider the mixture (with equal weights) of two deterministic local strategies such that $p(a,b,x,y)=\frac{1}{4}p(a,b \vert x,y)$ where for the first strategy we have $a=x$ and $b=0$ and for the second we have $a=x\oplus 1$ and $b=y$. As shown in the Table \ref{tab1} below (using a matrix representation) under the postselection $y=a$ the first strategy is mapped to a distribution such that $p(x,y)=1/2$ if $x\oplus y=0$ ($0$ otherwise) and the second is mapped to a distribution such that $p(x,y)=1/2$ if $x\oplus y=1$ ($0$ otherwise). That is, this post-selection of data generates measurement dependence between the hidden variable (choosing which deterministic strategy to use) and the measurement choices. 

Using the two mentioned local deterministic strategies and after the post-selection, we have a distribution $p(a,b,x,y)=\frac{1}{4}p(a,b \vert x,y)=\frac{1}{4}\delta_{a,y}\delta_{b,y(x \oplus 1)}$ that achieves the maximum violation the CHSH inequality ($=4$) and clearly is signalling.

\begin{table}[hp]
\begin{tabular}{c | c || c c | c c |}
$x\backslash y$ & & \multicolumn{2}{c|}{0} & \multicolumn{2}{c|}{1} \\
\hline
 & $a\backslash b$      &   0            &   1               &   0   &   1   \\
\hline
\hline
\multirow{3}{*}{0} & 0  &   $\frac{1}{4}$                      &   0            &       $\frac{1}{4}$        &      0        \\
                   & 1  &   0                      &   0            &       0        &      0        \\
\hline 
\multirow{2}{*}{1} & 0  &   0               &   0            &       0        &      0        \\
                   & 1  &   $\frac{1}{4}$                      & 0  &       $\frac{1}{4}$        & 0 \\
\hline
\end{tabular}
\quad
$\xrightarrow[\text{}]{\text{y=a}}$
\quad
\begin{tabular}{c | c || c c | c c |}
$x\backslash y$ & & \multicolumn{2}{c|}{0} & \multicolumn{2}{c|}{1} \\
\hline
 & $a\backslash b$      &   0            &   1               &   0   &   1   \\
\hline
\hline
\multirow{3}{*}{0} & 0  &   $\frac{1}{2}$                      &   0            &       0        &      0        \\
                   & 1  &   0                      &   0            &       0        &      0        \\
\hline 
\multirow{2}{*}{1} & 0  &   0               &   0            &       0        &      0        \\
                   & 1  &   0                      & 0  &       $\frac{1}{2}$        & 0 \\
\hline
\end{tabular}
\quad
\quad
,
\quad
\quad
\begin{tabular}{c | c || c c | c c |}
$x\backslash y$ & & \multicolumn{2}{c|}{0} & \multicolumn{2}{c|}{1} \\
\hline
 & $a\backslash b$      &   0            &   1               &   0   &   1   \\
\hline
\hline
\multirow{3}{*}{0} & 0  &    0                     &   0            &       0        &      0        \\
                   & 1  &   $\frac{1}{4}$                     &   0            &       0        &      $\frac{1}{4}$        \\
\hline 
\multirow{2}{*}{1} & 0  &   $\frac{1}{4}$               &   0            &       0        &      $\frac{1}{4}$        \\
                   & 1  &   0                      & 0  &       0        & 0 \\
\hline
\end{tabular}
\quad
$\xrightarrow[\text{}]{\text{y=a}}$
\quad
\begin{tabular}{c | c || c c | c c |}
$x\backslash y$ & & \multicolumn{2}{c|}{0} & \multicolumn{2}{c|}{1} \\
\hline
 & $a\backslash b$      &   0            &   1               &   0   &   1   \\
\hline
\hline
\multirow{3}{*}{0} & 0  &   0                      &   0            &       0        &      0        \\
                   & 1  &   0                      &   0            &       0        &      $\frac{1}{2}$        \\
\hline 
\multirow{2}{*}{1} & 0  &   $\frac{1}{2}$               &   0            &       0        &      0        \\
                   & 1  &   0                      & 0  &       0        & 0 \\
\hline
\end{tabular}
\caption{The two upper matrices show the matrix representation for the probability distribution $p(a,b,x,y)=\frac{1}{4}p(a,b \vert x,y)$ of the local deterministic strategy where $a=x$ and $b=0$ and how it transforms after performing the post-selection where $y=a$. The two lower matrices show the same for the deterministic strategy $a=x\oplus 1$ and $b=y$.}\label{tab1}
\end{table}

\subsection{Addressing the $X \rightarrow B$ loophole}

Differently from a usual Bell scenario, the instrumental causal structure is not subjected to the locality loophole since the measurement outcome $A$ has a direct causal influence over outcome $B$ also implying that in general $X$ and $B$ will be correlated. As we explain below, the instrumental scenario does introduces a new kind of loophole, which can in principle be addressed relying on interventions. Before discussing the loopholes in the instrumental scenario, first we draw parallels with the usual loopholes and meaning of device-independence in a usual Bell scenario.

The term device-independence refers to the fact that we can infer properties of our system of interest without relying on any assumptions about the internal mechanism of our measurement devices. However, still one has to rely on some global assumptions, external to the inner working of the devices. From a causal networks perspective (from which Bell's theorem is a particular case), device-independence refers to the fact once we impose the external mechanisms (the causal relations between the variables), our conclusions are independent of whatever are the internal mechanisms generating the value of a given variable (linear or non-linear response functions, Gaussian or non-Gaussian distributions, etc). 

For example, in a usual Bell scenario we associate the violation of a Bell inequality with a device-independent proof that the underlying quantum state is entangled (thus independent of the fact that we have measured a given observable, the dimension of the quantum state, etc). However, to achieve this conclusion we have to impose the locality and measurement independence assumptions. The question is then how can we be sure to fulfill such assumptions in an experiment?

To close the locality loophole we can rely on physical arguments since by invoking special relativity we can assure the locality assumption by measuring particles that are space-like separated. However, the presence of a possible measurement dependence (correlating the choice of observables and the source preparing the system to be measured) cannot be ruled out on a physical ground, thus introducing a loophole that cannot be closed. Clearly, however, measurement independence is an assumption external to the measurement devices. From the perspective of applications (such as cryptography) device-indepedence means that we can trust our results even without knowing our measurement apparatus that can then be treated as a black box. However, we do have to trust that the causal assumptions are being fulfilled otherwise an eavesdropper can easily simulate apparently quantum non-local correlations with classical resources (for example correlating the source and the measurement choices as has been done above).

The same is true for the instrumental scenario. As opposed to the locality loophole in Bell's theorem, the direct causal influence between the variables $X$ and $B$ (not mediated by variable $A$) cannot be ruled out by physical grounds such as special relatively. However, just as in the Bell case, once an instrumental inequality is violated we do not rely on any assumption about the measurement apparatus but only on external/causal assumptions. For example, in a clinical trial, the variable $X$ would stand for the treatment assigned to a patient (a drug or placebo), $A$ would represent the compliance of the patient (taking or not the treatment) and $B$ would be treatment response. As discussed in the main text, in such a randomized experiment, it is natural to assume that $X$ has not direct influence over $B$, however, if the patients discover that they are actually receiving placebo treatment, this might not be true anymore. That is, this loophole cannot be conclusively closed (unless interventions are made, see below), the best one can do is to design the best possible experiment (e.g, patients are unaware of the received treatment). Something similar happens with the measurement independence assumption in a Bell test. The best one can do is to improve our experimental trust in the measurement independence, for instance, using cosmic photons \cite{Handsteiner2017} or human randomness \cite{BBT2017}.  

Finally, as a potential way of ruling out such loophole one can make use of interventions. Notice that in the instrumental scenario the variables $X$ and $B$ are in general correlated but all these correlations are mediated by the variable $A$. This means that by intervening on $A$ (thus breaking the incoming causal link from $X$) the variable $X$ and $B$ should become independent, that is, $p(b \vert x, \mathrm{do}(a))=p(b\vert \mathrm{do}(a))$. However, if under interventions on $A$ we still observe correlations between $X$ and $B$, this could only be due some direct influence of $X$ into $B$ thus allowing us to detect such causal link. Of course that interventions are device-dependent operations but at least they provide an experimental way to addressing such loopholes.

To illustrate we give a very simple example. Suppose we have a probability distribution achieving the maximum algebraic value of \eqref{new_instrumental_SI} given by $\mathcal{I}_{\mathrm{max}}=7$. In this case, clearly we need a the direct causal influence $X \rightarrow B$. One way of achieving this value is via a deterministic strategy such that $a=0 \quad \forall x$ and $b=1$ if $x=1$ and $b=0$ if $x=2,3$ (thus $A$ has in fact no causal influence over $B$). One can easily check that $p(b \vert x, \mathrm{do}(a)) \neq p(b\vert \mathrm{do}(a))$ allowing one to conclude that some direct causal influence between $X$ and $B$ is present.

\subsection{The effect of detection inefficiencies}

As it happens with Bell tests, the experimental implementation of the violation of the instrumental\ inequality is also subjected to experimental imperfections. Here we analyze in details the effects of detections inefficiencies, that is, the fact that some of the generated photons might not lead to a detection click.

In our setup two entangled photons are generated and then measured in a projective basis, that is, defined the measurement direction we have two possible outcomes: $0$ (corresponding to the eigenvalue $+ 1$) and $1$ (corresponding to the outcome $- 1$). In this situation, the non-detection of a photon, labelled by $\emptyset$, can be handled by two different approaches.

First, one can treat the no-click event as third possible outcome of the performed measurement. However, notice that the instrumental inequality \eqref{new_instrumental_SI} refers to a scenario where only two outcomes are allowed. Thus, to treat the no-click as third event we have derive new instrumental inequalities taking that into account.

The second possibility, the one we follow here, is to remain in a description with two outcomes only and thus treat the no-click by a binning approach, that is, whenever a non-click happens we simply label $\emptyset$ as either $0$ or $1$. In this case we can simply use inequality \eqref{new_instrumental_SI}.

Quite generally a measurement with two outcomes can be represented by the POVM operators \cite{Chaves2011}
\begin{eqnarray}
\label{eq:POVM}
M_{\uparrow}= \eta_{\uparrow} \ket{\uparrow}\bra{\uparrow} +(1-\eta_{\downarrow}) \ket{\downarrow}\bra{\downarrow}, \\ \nonumber
M_{\downarrow}= \eta_{\downarrow} \ket{\downarrow}\bra{\downarrow} +(1-\eta_{\uparrow}) \ket{\uparrow}\bra{\uparrow},
\end{eqnarray}
where $\eta_{\uparrow}=\eta_{\downarrow}=1$ represent the case of perfect projective measurements and where $\ket{\uparrow,\downarrow}$ represent some arbitrary basis for qubits.

\begin{figure}[t!]
\center
\includegraphics[width=1\columnwidth]{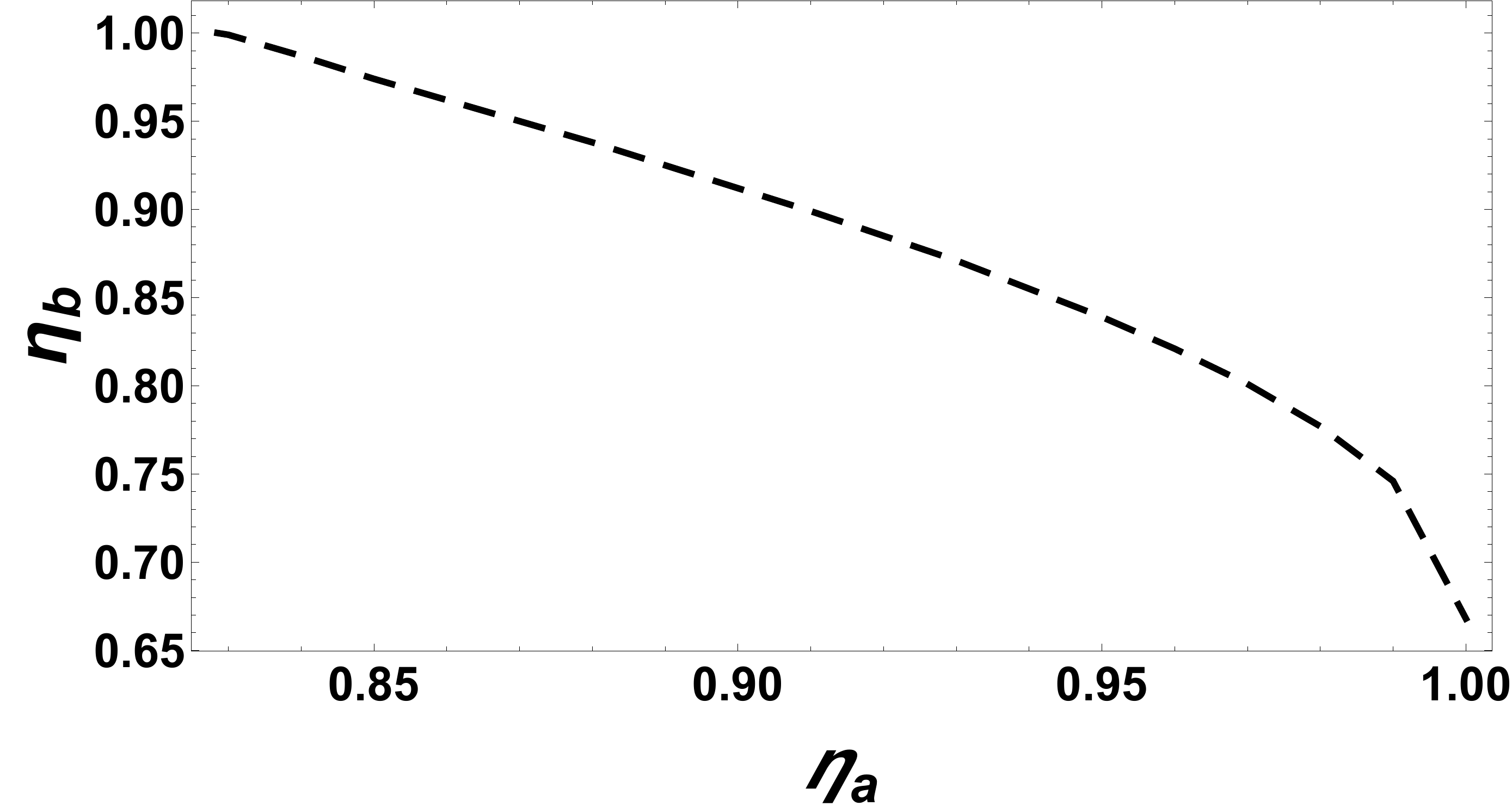}
\caption{\textbf{Critical values above which one can violate the instrumental inequality \eqref{new_instrumental_SI}:} At  $(\eta_a,\eta_b)=(0.8285,1)$ the optimal states are maximally entangled state while at $(\eta_a,\eta_b)=(1,0.667)$ we need an almost separable state.}
\label{fig:etaplot}
\end{figure}

To model the binning of the no-click event we can make $\eta_{\uparrow}=1$ and $\eta_{\downarrow}=\eta$ (where $\eta$ is the detection efficiency of the photon detector). This corresponds to the case where whenever we obtain $\emptyset$  we simply relabel it as the outcome $0$ (corresponding to projection in $\uparrow$). In other terms the event $\emptyset$ is binned with the event $0$. Using this modeling and optimizing over the entangled states and measurement basis we have obtained the minimum detection efficiencies required to obtain a detection-loophole-free violation of inequality \eqref{new_instrumental_SI}. The results are shown in Fig. \ref{fig:etaplot}.

We observe that there is an asymmetry between the detection efficiencies of the $\eta_a$ and $\eta_b$ corresponding to the measurements outcomes $A$ and $B$, respectively. Setting $\eta_a=1$ we obtain the critical value $\eta^{\mathrm{crit}}_b \approx 0.667$, below which no violation of the instrumental inequality is possible anymore. In turn, for $\eta_b=1$ we observe $\eta^{\mathrm{crit}}_a \approx 0.828$. That is, the violation is more robust according to the detection inefficiencies of the measurement corresponding to the $B$ variable. Instead, if we set $\eta_a=\eta_b=\eta$ we obtain $\eta^{\mathrm{crit}} \approx 0.905$.

Further, we observe that for detection inefficiencies in $a$, the optimal states are always maximally entangled. In turn, for detection inefficiencies in $b$, the best violations are obtained by states with reduced entanglement that in fact tend to separable states as $\eta_b \rightarrow \eta^{\mathrm{crit}}_b$. This is a similar effect to that in usual Bell tests \cite{Eberhard1993}, such that the best critical detection efficiencies are obtained for states close to separability.

\subsection{The effects of other sources of experimental errors}

As shown in the main text, the expected quantum violation of the instrumental inequality is  $ \mathcal{I}_{\mathrm{Q}}=1+2\sqrt{2}$, achieved by a maximally entangled state of two qubits. In our experimental photonic setup, in order to prepare one of those states it is necessary to use a Spontaneous Parametric Down Conversion (SPDC)-type II source. It has been shown \cite{Noise_Cabello_1}\cite{Noise_Cabello_3} that these sources suffer from two different kinds of noise: white noise and colored noise. Thus, the experimental state given by the mixture of these two noises can be modeled as:
\begin{eqnarray}
\label{eq:noise_source}
\rho_{noise}= & & v\ket{\phi^{+}}\bra{\phi^{+}}+ \\ \nonumber
& & +(1-v)\left[\lambda\frac{\ket{\phi^{+}}\bra{\phi^{+}}+ \ket{\phi^{-}}\bra{\phi^{-}}}{2}+(1-\lambda)\frac{\mathbb{I}}{4}\right].
\end{eqnarray}
White noise is responsible for the appearance of the maximally mixed state and coloured noise for the appearance of $\ket{\phi^{-}}\bra{\phi^{-}}$.  

\begin{figure}[h!]
\center
\includegraphics[width=1\columnwidth]{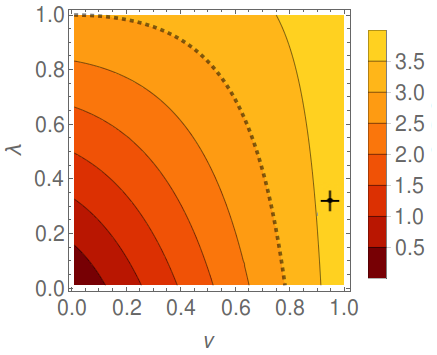}
\caption{\textbf{EPR source noise effect on inequality parameter:} In this graph is represented, by a scale of colors as showed in the label, the maximum values of $\mathcal{I}_{\mathrm{Q}}$ for any $v~\text{and}~\lambda$ according to the state in \eqref{eq:noise_source}. The black cross denotes the estimated working point of our EPR source while the dashed line separates the region of violations (to the right) from the region of no-violation of the instrumental inequality (to the left).}
\label{fig:Noise_Graphic}
\end{figure}

Considering this state and optimizing over the measurement settings for each value of $v~\text{and}~\lambda$ we obtain the results shown in Fig. \ref{fig:Noise_Graphic}. The point marked in this figure represents the experimental violation expected from  the characterization of our EPR source that corresponds to $v\approx0.94$ and $\lambda\approx 0.33$ yielding $\mathcal{I}_{\mathrm{Q}}=3.643$. Since half wave-plates and fibers don't introduce significant noise we can consider this point as the theoretical expected violation for the experiment based on post-selection of data. Therefore, it's clear that our experimental result of $\mathcal{I}_{\mathrm{Q}}=3.621 \pm 0.023$ is in almost perfect agreement with the theoretical one.

\begin{figure}[h!]
\center
\includegraphics[width=1\columnwidth]{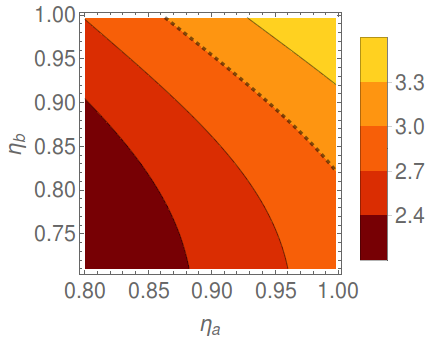}
\caption{\textbf{No-click and EPR source noise effect on inequality parameter:} Maximum violations reached in a no-click scenario by maximizing over the state and measurement settings considering the presence of colored and white noise at the working point of our EPR source. The magnitude of violation is represented by a scale of colors separated by contours. The dashed line separates the region where a violation occurs (on the right) from the region where $\mathcal{I}_Q$ is lower than the classical bound (on the left).}
\label{fig:Noise_Graphic_NoClick}
\end{figure}

In order to confirm also the Pockels Cell experimental results, we need to model the non-perfect efficiency of the crystal implementing $\sigma_{z}$. We thus have to take into account the possibility in which the Pockels Cell acts as identity instead of $\sigma_{z}$. This can be modeled by a Pockels Cell visibility $v_{pc}$ that can be interpreted as the probability of not applying the $\sigma_{z}$ transformation. Thus, the case when $a=0$ (the case when the $\sigma_{z}$ should be applied) can be computed as 
\begin{equation}
\label{eq:noise_pc}
p_{\mathrm{Q}}(0,b|x)=\Tr\left[\left(M^{x}_{0}\otimes M^{1}_{b}\right)\rho^{pc}_{noise}\right],
\end{equation}
where
\begin{equation}
\rho^{pc}_{noise}=(1-v_{pc})\rho_{noise}+v_{pc}\mathbb{I}^{A}\otimes\sigma^{B}_{z}\rho_{noise}\mathbb{I}^{A}\otimes\sigma^{B}_{z}.
\end{equation}
All other probabilities should be computed just as in equation \eqref{PQ} but considering a state given by $\rho_{noise}$. We have evaluated experimentally the crystal visibility which was found to be $v_{pc}=0.87$. Together with the other sources of errors we find the expected violation to be $\mathcal{I}_{\mathrm{Q}}=3.342$ which is in very good agreement with our experimental result of   $\mathcal{I}_{\mathrm{Q}}=3.358 \pm 0.020$. 
We want to highlight that the previous result can be achieved also considering a POVM (Positive operator valued measure) instead of modifying the probability as done in \eqref{eq:noise_pc}. To show it, we can define such POVM by the following operators:
\begin{eqnarray}
&&E_0^0=v_{pc}\ket{\tilde{0}}\bra{\tilde{0}} + (1-v_{pc})\ket{\slashed{0}}\bra{\slashed{0}}  \label{eq:povm1}\\
&&E_1^0=v_{pc}\ket{\tilde{1}}\bra{\tilde{1}} + (1-v_{pc})\ket{\slashed{1}}\bra{\slashed{1}} \label{eq:povm2}
\end{eqnarray}
where $\ket{\slashed{0}},\ket{\slashed{1}}$ are the eigenvectors of $\frac{\sigma_z-\sigma_x}{\sqrt{2}}$ while  $\ket{\tilde{0}},\ket{\tilde{1}}$ represent the eigenspace of $\frac{\sigma_z+\sigma_x}{\sqrt{2}}$. Replacing the projective measurements $M^0_b$ with eq. \eqref{eq:povm1} and \eqref{eq:povm2}, we obtain again $\mathcal{I}_Q =3.342$.

Now, we consider how the no-click events joined with EPR source noise (considered in the post-selection experiment) would affect  the maximal violation achievable. To take into account no-click events and the possibility that the maximum value of $\mathcal{I}_{\mathrm{Q}}$ is not always reached by maximally entangled state for all values of $\eta_{a}~  \textnormal{and}~\eta_{b}$, we modified the state in eq. \eqref{eq:noise_source} as following:
\begin{eqnarray}
& &\ket{\phi_{\theta}}\equiv\cos{\theta}\ket{\uparrow\uparrow}+\sin{\theta}\ket{\downarrow\downarrow}\\
& &\ket{\phi_{\theta}^{\bot}}\equiv\sin{\theta}\ket{\uparrow\uparrow}-\cos{\theta}\ket{\downarrow\downarrow}
\end{eqnarray}
\begin{eqnarray}
\rho_{noise}= & & v\ket{\phi_{\theta}}\bra{\phi_{\theta}}+ \\ \nonumber
& & +(1-v)[\lambda\frac{\ket{\phi_{\theta}}\bra{\phi_{\theta}}+ \ket{\phi_{\theta}^{\bot}}\bra{\phi_{\theta}^{\bot}}}{2}+(1-\lambda)\frac{\mathbb{I}}{4}]
\end{eqnarray}
Fixing the values of $\lambda$ and $v$ ($0.33$ and $0.94$ respectively), maximizing the violation over measurement settings and $\theta$ in the POVM scenario depicted in eq. \eqref{eq:POVM} with $\eta_{\uparrow}^{a,b}=1$ and $\eta_{\downarrow}^{a,b}=\eta_{a,b}$,  contours in Fig. \ref{fig:Noise_Graphic_NoClick} were obtained.

Since the parameters $\lambda$ and $v$, reported above, were estimeted exploiting the accidental counts correction method, all the previous analyzes are valid only in the case we compare our theoretical expectation values with the experimental ones where such correction was taken into account. Therefore, in order to perform a similar analysis to predict the violation values without correction, we have to just modify the state $\rho_{noise}$ adding a further term of white noise:
\begin{eqnarray}
\rho_{noise}^{noacc} = &&\gamma \biggl[ v\ket{\phi^+}\bra{\phi^+}+ \biggr. \\ 
&&\biggl.+(1-v) \left( \lambda\frac{\ket{\phi^+}\bra{\phi^+}+\ket{\phi^-}\bra{\phi^-}}{2}+(1-\lambda) \frac{\mathbb{I}}{4}\right) \biggr]+ \nonumber \\ 
&&+(1-\gamma)\frac{\mathbb{I}}{4}. \nonumber
\end{eqnarray}
where $\gamma$ is a parameter which can be experimentally calculated as follows:
\begin{equation}\label{eq:gamma}
\gamma=1 - \frac{\sum_{a,b}Acc_{a,b}}{\sum_{a,b}Coinc_{a,b}}
\end{equation}
in Eq. \eqref{eq:gamma} $Acc_{a,b}$ and $Coinc_{a,b}$ are, respectively, the accindental counts and the total coincidence counts observed by detectors associated to the a, b outcomes. From our data we have obtained $\gamma \approx 0.971$ which leads to the following theoretical expectation values:
\begin{eqnarray}
&&\mathcal{I}_Q = 3.537,~~~\textnormal{for the post-selection experiment.}\nonumber \\
&&\mathcal{I}_Q = 3.245,~~~\textnormal{for the active feed-forward experiment.}\nonumber
\end{eqnarray}
Those results are again in agreement with the experimental values reported in the main text.